\documentclass[preprintnumbers,amsmath,amssymb, prd,superscriptaddress]{revtex4}
\usepackage{latexsym}
\usepackage{amssymb}
\usepackage{epsfig,amsmath,graphics}
\usepackage{epstopdf}
\usepackage{verbatim}
\usepackage[hypertex]{hyperref}
\usepackage{graphicx}
\usepackage{subfigure}
\usepackage{verbatim}
\usepackage{ulem}
\usepackage{dsfont}

\newcommand{\OO}{\mathcal{O}}

\newcommand{\GeV}{\text{GeV}}
\newcommand{\TeV}{\text{TeV}}

\newcommand{\s}{\text{s}}

\newcommand\ee{\end{equation}}
\newcommand\be{\begin{equation}}

\newcommand{\LiSeven}{^{7}\text{Li}}
\newcommand{\LiSix}{^{6}\text{Li}}
\newcommand{\FiveBar}{\bar{5}}
\newcommand{\TenBar}{\overline{10}}

\newcommand{\FiveBarDagger}{\bar{5}^{\dagger}}
\newcommand{\TenDagger}{10^{\dagger}}

\newcommand{\HuDagger}{H_{u}^{\dagger}}
\newcommand{\huDagger}{h_{u}^{\dagger}}

\newcommand{\OrderOne}{\mathcal{O}\left(1\right)}
\newcommand{\lambdac}{\lambda_{ij}}
\newcommand{\FortyFiveBar}{\overline{45}}

\newcommand{\Mgut}{M_\text{GUT}}
\newcommand{\gut}{\text{GUT}}

\newcommand{\huvev}{\langle h_u \rangle}
\newcommand{\hdvev}{\langle h_d \rangle}
\newcommand{\phibardoublet}{\phibar^{(2)} }
\newcommand{\phidoublet}{ \phi^{(2)}}
\newcommand{\phibardoubletvev}{\langle \phibar^{(2)} \rangle}
\newcommand{\phidoubletvev}{\langle \phi^{(2)} \rangle}
\newcommand{\tildephidoublet}{\tilde{\phi}^{(2)}}
\newcommand{\tildephibardoublet}{\tilde{\phibar}^{(2)}}

\newcommand{\loopfactor}{\frac{1}{16 \pi^2}}

\newcommand{\doubletbmu}{B_{\mu}^{(2)}}
\newcommand{\tripletbmu}{B_{\mu}^{(3)}}
\newcommand{\hutriplet}{h_{u}^{(3)}}
\newcommand{\hdtriplet}{h_{d}^{(3)}}
\newcommand{\FSUSY}{\langle F \rangle}
\newcommand{\sigmavev}{\langle\sigma\rangle}
\newcommand{\Sigmavev}{\langle\Sigma\rangle}

\newcommand{\mintwoloopten}{-4 \, \epsilon_2 }

\newcommand{\minfourloopten}{\epsilon_2 ^2}

\newcommand{\phibar}{\overline{\phi}}

\def\r{\right)}
\def\l{\left(}

\begin{document}


\title{A Domino Theory of Flavor}

\author{Peter W. Graham}
\affiliation{Department of Physics, Stanford University, Stanford, California 94305}
\affiliation{SLAC National Accelerator Laboratory, Stanford University, Menlo Park, California 94025}

\author{Surjeet Rajendran}
\affiliation{Department of Physics, Stanford University, Stanford, California 94305}
\affiliation{SLAC National Accelerator Laboratory, Stanford University, Menlo Park, California 94025}

\preprint{SLAC-PUB-13734}

\date{\today}

\begin{abstract}
We argue that the fermion masses and mixings are organized in a specific pattern.  The approximately equal hierarchies between successive generations, the sizes of the mixing angles, the heaviness of just the top quark, and the approximate down-lepton equality can all be accommodated by many flavor models but can appear ad hoc.  We present a simple, predictive mechanism to explain these patterns.  All generations are treated democratically and the flavor symmetries are broken collectively by only two allowed couplings in flavor-space, a vector and matrix, with arbitrary $\OO(1)$ entries.  Repeated use of these flavor symmetry breaking spurions radiatively generates the Yukawa couplings with a natural hierarchy.  We demonstrate this idea with two models in a split supersymmetric grand unified framework, with minimal additional particle content at the unification scale.  Although flavor is generated at the GUT scale, there are several potentially testable predictions.  In our minimal model the usual prediction of exact b-$\tau$ unification is replaced by the SU(5) breaking relation $m_\tau / m_b = 3 / 2$, in better agreement with observations.  Other SU(5) breaking effects in the fermion masses can easily arise directly from the flavor model itself.  The symmetry breaking that triggers the generation of flavor necessarily gives rise to an axion, solving the strong CP problem.  These theories contain long-lived particles whose decays could give striking signatures at the LHC and may solve the primordial Lithium problems.  These models also give novel proton decay signatures which can be probed by the next generation of experiments.  Measurement of the various proton decay channels directly probes the flavor symmetry breaking couplings.  In this scenario the Higgs mass is predicted to lie in a range near 150 GeV.
\end{abstract}

\maketitle

\tableofcontents

\section{Flavor Philosophy}

The Yukawa couplings of the Standard Model (SM) are not random $\OO(1)$ numbers.  This could be called a failure of the effective field theory approach, or equivalently, evidence for new physics above the weak scale.  Our philosophy is that the mere smallness of the Yukawas is not as important as their nontrivial structure.  Even on a log scale they do not appear to be randomly distributed, as they are all $\gtrsim 10^{-5}$.  Further, they are evenly spaced.  Within each type of fermion (up-type quarks, down-type quarks, or charged leptons) the masses of successive generations are always split by about two orders of magnitude.  Finally, the down-type quark and lepton masses of each generation are similar, and different from the up-type quark masses, suggesting a grand unified theory (GUT).  While it is possible that such patterns happen to result from an underlying theory that acts in an independent or random way on each fermion, we will follow the philosophy that these patterns are the result of a single, unified structure acting in a universal, flavor-blind way on all the quarks and leptons.

Several other general approaches to flavor have been proposed.  Froggatt-Nielsen type models have perhaps the most similarity with our mechanism but usually require careful choices of new charges for the different SM quarks and leptons as well as many new fields in order to accommodate the observed hierarchical structure \cite{Froggatt:1978nt, Froggatt:1979sz, Buchmuller:2006ik, Lebedev:2007hv, Randall:2009dw}.  Models with family or horizontal flavor symmetries can also generate the Yukawa couplings but with complicated new degrees of freedom, symmetries, flavor breaking mechanisms, and arranged choices of charges for the different fermions \cite{Liu:1996ua, Leurer:1993gy, King:2003rf, Appelquist:2006ag, Ko:2007dz, Lykken:2008bw}.  Models without imposed flavor symmetries still frequently have new structures \cite{Ferretti:2006df}.
Extra-dimensions have also been invoked to explain the smallness of the Yukawa couplings \cite{Hall:2001rz, Hebecker:2002re, ArkaniHamed:1999dc, Agashe:2008fe, Perez:2008ee}.  These often involve choosing different positions to localize the different quark and lepton fields of each generation.  Models exploiting exponentially suppressed wavefunction overlaps can generically produce much smaller Yukawa couplings than observed (in fact even Yukawas of the size necessary for Dirac neutrinos \cite{Agashe:2008fe}), and can produce any pattern of masses distributed randomly on a logarthmic scale.
These frameworks can explain why the Yukawa couplings are small numbers, but do not as directly explain the observed hierarchical pattern since they could just as easily accommodate any other pattern of small numbers.  Such models are less constrained and thus, in that sense, less predictive than our model.

We will generate the flavor structure of the SM from a theory that treats all generations identically.  We work in a unified theory and add one new symmetry to forbid the Yukawas, in its simplest form just a global U(1) or discrete group, with simple charge assignments that do not distinguish between generations (e.g. see Table \ref{Tab:charges}).  We also add one new singlet and one new vector-like pair of fields and write down all allowed operators with $\OO(1)$ coefficients.  This generates just two couplings with flavor indices, i.e.~two directions in `flavor-space'.  When the symmetry is spontaneously broken, all the Yukawa couplings are generated radiatively.  But, because of the simple structure of the two allowed flavor directions, the masses of successive generations arise at successive orders.  This gives rise to the observed hierarchical pattern of fermion masses and mixings without the need for making different choices (such as of charge or extra-dimensional localization) for the different fermions.

Radiative models of flavor have a long history \cite{Weinberg:1972ws, Georgi:1972mc, Georgi:1972hy, He:1989er, Babu:1990vx, Hashimoto:2009xi}.  Several of the models have some similarities to the flavor structure of our model \cite{Balakrishna:1987qd, Balakrishna:1988ks, Banks:1987iu}, and especially \cite{Dobrescu:2008sz}, including some with supersymmetry (SUSY) \cite{ArkaniHamed:1995fq, ArkaniHamed:1996zw, Babu:1989tv, Nandi:2008zw} or in a GUT \cite{Babu:1989tv, Ibanez:1981nw, Barr:2008kg, Barr:2007ma, Babu:1989fg}.
Many of these models only generate some but not all of the Yukawa couplings or require complicated new sectors.  Additionally, many low scale flavor models are now in conflict with experimental flavor constraints.

\section{General Domino Framework}

Our goal is to explain the hierarchical pattern of fermion masses and mixing angles, as opposed to merely accommodating the observed smallness of the Yukawa couplings.  In this Section we give the basic idea behind our models.  Although we generate the Yukawas from radiative corrections, the mechanism responsible for the pattern of hierarchies is more general and may well find application in other models.

\subsection{The Model}
\label{DominoPhilosophy}

To simplify the discussion, we consider here just the couplings in our theory which carry flavor information.  These `flavorful' parts of the superpotential are
\begin{equation}
\label{Eqn: flavor superpot}
\mathcal{W} \supset y_3 \, 10_3 \, 10_3 \, H_u  + \lambda_{ij} \, 10_i \, \FiveBar_j \, \phibar
\end{equation}
where $10_i$ is the ith generation of $10$, $H_u$ the light higgs whose vev breaks electroweak symmetry, $y_3$ the top yukawa and $\lambda$ is a random matrix with $\OrderOne$ entries.  Currently, $\phibar$ is a new field that does not get a vev, though we will see in Section \ref{Model5} that it can just be the down-type higgs, $H_d$.

The top Yukawa term (the first term in Eqns. \eqref{Eqn: flavor superpot}) might appear to violate our philosophy since it appears to require a specific choice of direction in flavor space.  However, we can rewrite it in a more suggestive, basis-independent way as
\begin{equation}
\label{Eqn: spurion superpot}
\mathcal{W} \supset \left(c_i \, 10_i  \right) \, \left( c_j \, 10_j \right) \, H_u  + \lambda_{ij} \, 10_i \, \FiveBar_j \, \phibar
\end{equation}
where $c_i$ is a random vector with $\OO(1)$ entries.  Of course we can always choose a basis in which $c$ points entirely in the 3 direction (i.e. only $c_3$ is nonzero) which then gives Eqn. \eqref{Eqn: flavor superpot}.  The form of the superpotential in Eqn. \eqref{Eqn: spurion superpot} indicates how this term can be generated in a flavor-blind manner.  We will simply allow a term in the underlying model which has the arbitrary vector $c$ in `10-flavor space'.  Since $c$ is only a single vector in `10-space', it can only generate a Yukawa coupling in one direction in $10 \otimes 10$ space, as seen in Eqn. \eqref{Eqn: spurion superpot}.  By choice of basis, we can call this the $10_3 \otimes 10_3$ direction, or equivalently the top Yukawa coupling.  This gives only the top quark an $\OO(1)$ Yukawa coupling and hence a mass around the Higgs vev.  At this order, all other fermions are massless.  The details of such a model will be given in Section \ref{TopYukawa}.

\subsection{Hierarchical Masses}
\label{Spurions}

Note that the superpotential in Eqn. \eqref{Eqn: spurion superpot} breaks all the flavor symmetries $U(3)_{10} \times U(3)_{\FiveBar}$.  The top Yukawa breaks $U(3)_{10} \to U(2)_{10}$, and the $\lambda$ term breaks the rest of the symmetries.  Since the top quark is connected to the Higgs vev and all the flavor symmetries are broken, we expect all the Yukawa couplings to be generated by radiative corrections.  Of course, the main point of the model is to generate the hierarchical pattern of masses and mixings so we must avoid generating all the remaining Yukawa couplings at the same order.  Naively this might seem to be a problem in the simple model of Eqn. \eqref{Eqn: spurion superpot} since all the flavor symmetries are broken at the same level by arbitrary, flavor-blind couplings.  However, we can see from a simple spurion-type analysis that the Yukawa couplings are generated in the observed hierarchical pattern.

There are only two couplings in the model that carry flavor information, $c$ and $\lambda$, so all the Yukawa couplings must be made out of these.  Since $c$ only gives the top quark, to generate other Yukawas we must use $\lambda$'s.  Taking the combination $\lambda^\dagger c$ produces a vector in $\FiveBar$-flavor space, which will be the bottom and tau direction.  Note that at this order, only one direction in $\FiveBar$ space, i.e. one generation of downs and leptons, has been given a Yukawa coupling.  So all the downs and leptons besides the $b$, $\tau$ have not yet gotten a mass.  At the next order, $\lambda \lambda^\dagger c$ is the charm direction.  Again note that at this order we have only generated one more mass, and so we still have not generated a mass for the up-quark.  Thus the up sector Yukawa couplings, $y^u_{ij} \, H_u \, 10_i \, 10_j$, are generated at successive orders:
\begin{equation}
\label{Eqn: up-yukawa spurions}
\begin{array}{rclcrcl}
c \otimes c & \leadsto y^u_{33} = & \text{top mass} \propto 1 & \qquad &
c \otimes \left( \lambda \lambda^\dagger \right) c & \leadsto y^u_{32} = & \text{top-charm mixing} \propto \epsilon \\
\left( \lambda \lambda^\dagger \right) c \otimes \left( \lambda \lambda^\dagger \right) c & \leadsto y^u_{22} = & \text{charm mass} \propto \epsilon^2 & \qquad &
c \otimes \left( \lambda \lambda^\dagger \right)^2 c & \leadsto y^u_{31} = & \text{top-up mixing} \propto \epsilon^2 \\
\left( \lambda \lambda^\dagger \right)^2 c \otimes \left( \lambda \lambda^\dagger \right)^2 c & \leadsto y^u_{11} = & \text{up mass} \propto \epsilon^4 & \qquad &
\left( \lambda \lambda^\dagger \right) c \otimes \left( \lambda \lambda^\dagger \right)^2 c & \leadsto y^u_{21} = & \text{charm-up mixing} \propto \epsilon^3 \\
\end{array}
\end{equation}
where we have parametrized each insertion of $\lambda \lambda^\dagger$ with a factor of $\epsilon$.  In the specific models we present in the next sections we will calculate $\epsilon$ precisely.  It will arise because each insertion of $\lambda$ is a vertex with a $\phibar$ field coming out, as in Eqn. \eqref{Eqn: spurion superpot}.  These $\phibar$'s must then be contracted with each other, making each insertion of $\lambda \lambda^\dagger$ a loop factor.

The down quark Yukawas, $y^d_{ij} \, H^\dagger_u \, 10_i \, \FiveBar_j$, are similarly generated in a hierarchical pattern, and the masses are generally smaller than in the up sector:
\begin{equation}
\label{Eqn: down-yukawa spurions}
\begin{array}{rclcrcl}
c \otimes \lambda^\dagger c & \leadsto y^d_{33} = & b, \, \tau \text{ mass} \propto
\delta & \qquad &
c \otimes \left( \lambda^\dagger \lambda \right) \lambda^\dagger c & \leadsto y^d_{32} = & b-s \text{ mixing} \propto \delta \epsilon \\
\left( \lambda^\dagger \lambda \right) c \otimes \left( \lambda^\dagger \lambda \right) \lambda^\dagger c & \leadsto y^d_{22} = & s, \, \mu \text{ mass} \propto \delta \epsilon^2 & \qquad &
c \otimes \left( \lambda^\dagger \lambda \right)^2 \lambda^\dagger c & \leadsto y^d_{31} = & b-d \text{ mixing} \propto \delta \epsilon^2 \\
\left( \lambda^\dagger \lambda \right)^2 c \otimes \left( \lambda^\dagger \lambda \right)^2 \lambda^\dagger c & \leadsto y^d_{11} = & d, \, e \text{ mass} \propto \delta \epsilon^4 & \qquad &
\left( \lambda^\dagger \lambda \right) c \otimes \left( \lambda^\dagger \lambda \right)^2 \lambda^\dagger c & \leadsto y^d_{21} = & s-d \text{ mixing} \propto \delta \epsilon^3 \\
\end{array}
\end{equation}
where $\delta$ stands for the cost of inserting a single $\lambda^\dagger$ which we will calculate in our specific models.

The mixing angles in the CKM matrix will come from a combination of the off-diagonal terms in the up and down Yukawa matrices.  We will calculate these in detail later, but we can already see that these mixings are at roughly the square root of the mass ratios.  For example, the top-charm mixing arises at order $\epsilon$ while top-charm or bottom-strange mass ratio is at order $\epsilon^2$.  This is a famously noticed phenomenological relation.

\begin{figure}
\begin{center}
\subfigure[ ]{\label{Fig: observed masses}}
\includegraphics[width=2.5 in]{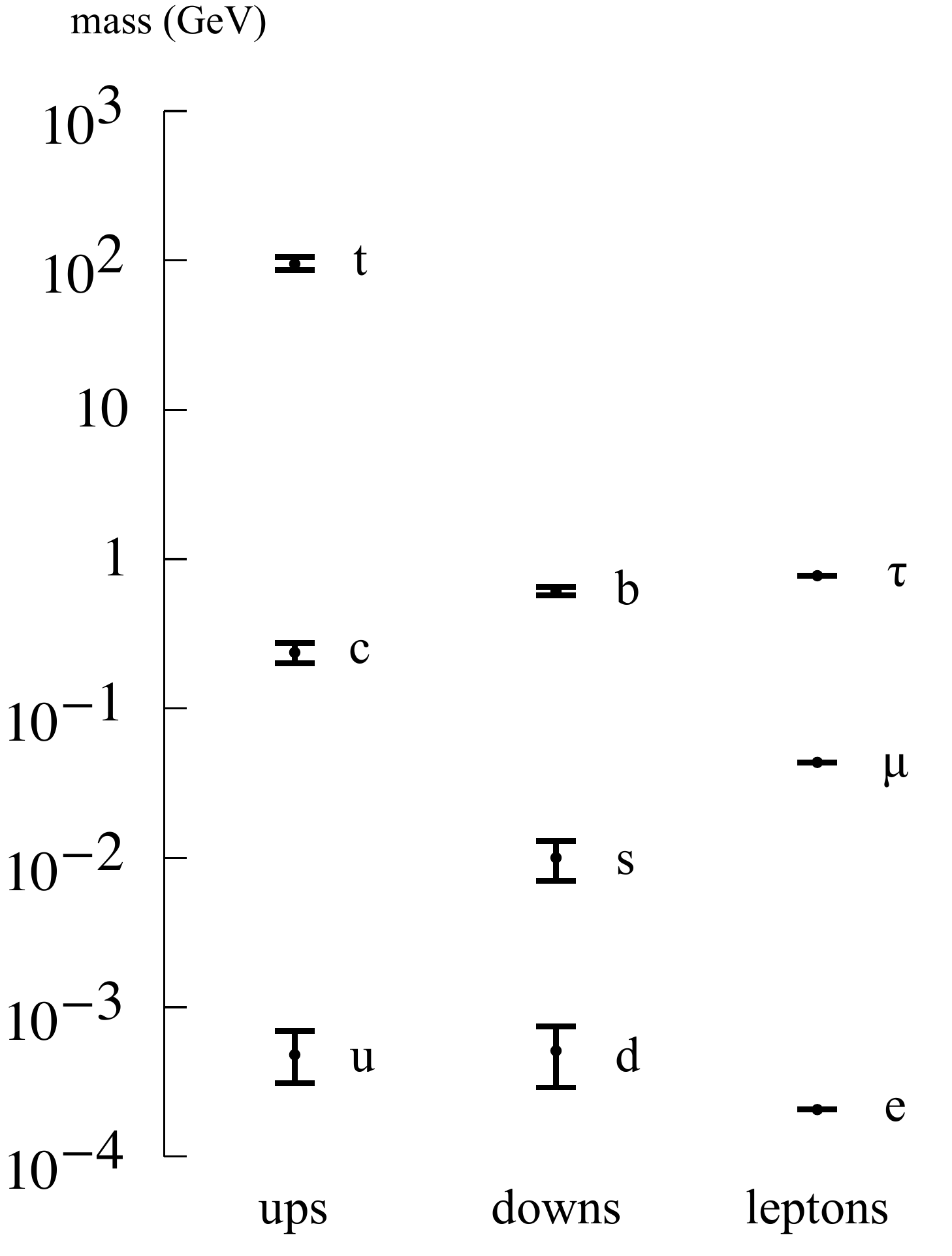} \qquad \qquad \qquad
\subfigure[ ]{\label{Fig: our pattern}}
\includegraphics[width=2 in]{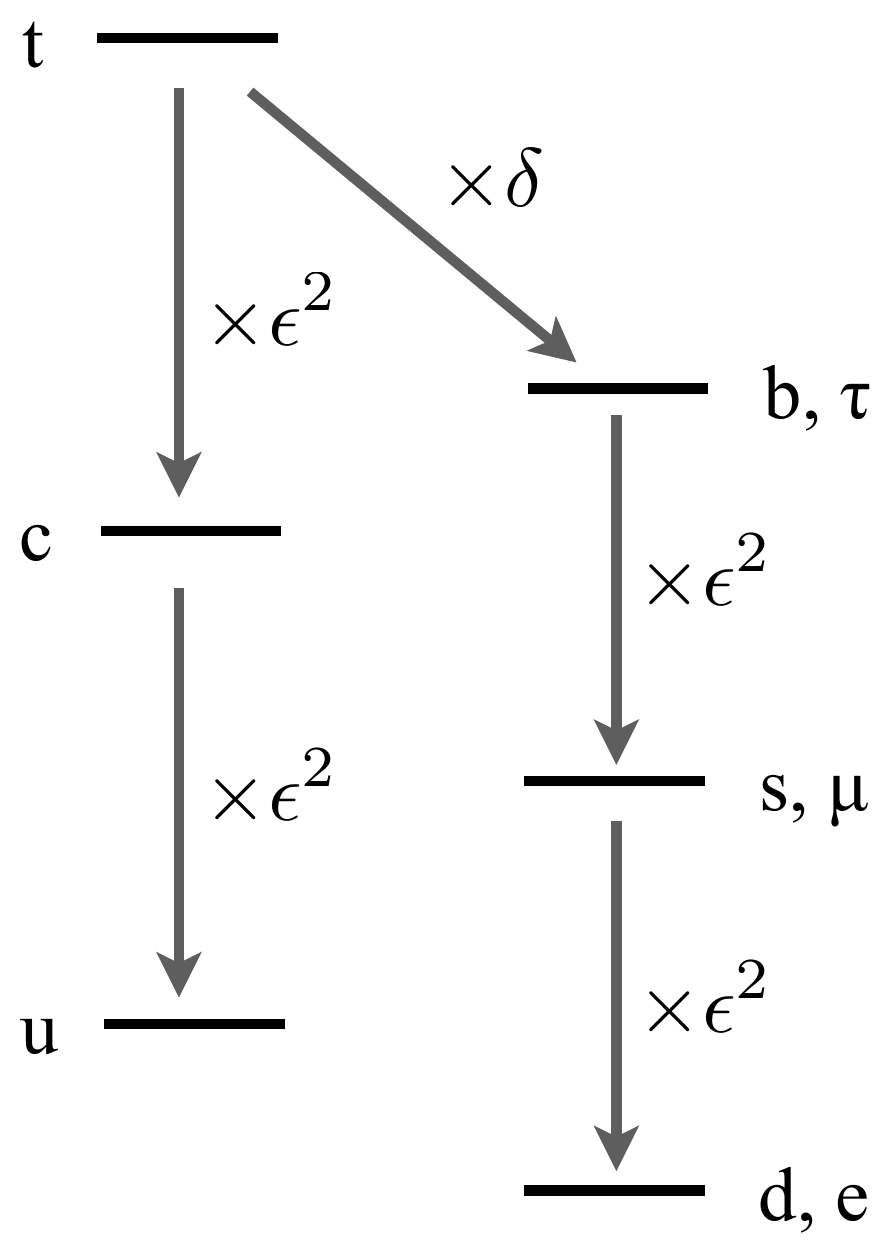}
\caption{ \label{Fig: dominopattern} Figure \ref{Fig: observed masses} shows the values of the Yukawa couplings at the GUT scale.  These are shown for the MSSM with large $\tan \beta$ as an example \cite{Xing:2007fb}.  The SM values are similar.  Figure \ref{Fig: our pattern} shows the hierarchical pattern of masses generated by our domino mechanism.  Each arrow represents a spurion factor of either $\delta$ for the bottom to top ratio, or $\epsilon^2$ for the others.}
\end{center}
\end{figure}

We follow the standard effective field theory philosophy that any allowed coupling should be present at $\OO(1)$.  Then, we simply allow two terms in the Lagrangian which carry flavor information, an arbitrary vector in `10-space' and an arbitrary matrix in `$10 \otimes \FiveBar$ space'.  These two flavor-symmetry breaking spurions then generate all the fermion masses and mixing angles in a hierarchical pattern as seen in Eqns.~\eqref{Eqn: up-yukawa spurions} and \eqref{Eqn: down-yukawa spurions} and in Fig.~\ref{Fig: dominopattern}.  The hierarchies between generations arise even though the original theory did not distinguish between any of the generations.  All of the original couplings are `flavor-blind' in that all three generations are treated identically.  They are not, for example, given different charges under a flavor symmetry, or located at different points in an extra-dimension.  It is just the structure of these two allowed couplings that causes the three generations to get such hierarchically split masses.  In fact, this mechanism is general and could possibly be implemented in a framework other than radiative generation.

\section{Models}
In this section, we construct simple models that embody the spirit of the GUT inspired domino mechanism. For concreteness, we work in the context of $SU(5)$ GUTs using the superpotential (\ref{Eqn: flavor superpot}). In (\ref{Eqn: flavor superpot}), gauge invariance requires $\phibar$ to be either a $\FiveBar$ or a $\FortyFiveBar$. While the general mechanism for generating flavor is similar for both representations, there are representation dependent details that arise in specific models. Accordingly, we discuss two models, classified by the representation of $\phibar$. The models are consistent with the general philosophy of the paper wherein we allow all terms allowed by symmetry, with $\OrderOne$ coupling constants and democratic treatment of all the generations. We first discuss the case where $\phibar$ is a $\FiveBar$ in section \ref{Model5}, where we discuss the complete flavor generation mechanism including the quark and lepton mass matrices, the CKM matrix and GUT breaking effects. In section \ref{Model45}, we consider the case where $\phibar$ is a $\FortyFiveBar$ and discuss the details of this model that are different from the $\FiveBar$ case.

\subsection{Minimal Model}
\label{Model5}

The down-type higgs field $H_d$ is a natural candidate for $\phibar$ when it is in the $\FiveBar$ representation of $SU(5)$. With $\phibar = H_d$, the superpotential (\ref{Eqn: flavor superpot}) becomes 
\begin{equation}
\mathcal{W} \supset y_3 \, 10_3 \, 10_3 \,  H_u  + \lambdac \,10_i \, \FiveBar_j \, H_d
\label{InitialSuperPotential}
\end{equation}
In the spirit of this paper, the couplings $y_3$ and $\lambdac$ are all $\OrderOne$. As we will see, from this starting point with no new fields beyond the normal content of the MSSM, the entire flavor structure of the SM is generated radiatively.

A vev $\hdvev$ for $h_d$ (the scalar, SU(2) doublet, component of the superfield $H_d$) makes comparable contributions to the masses of all the three generations. In the absence of fine tuning, this theory can reproduce the electron mass $m_e$ only if $\langle h_d \rangle \lessapprox m_e$. In order to get the correct top mass $m_t$, we must have $\huvev \sim m_t$. These constraints together imply that
\begin{equation}
\frac{1}{\tan \beta} = \frac{\hdvev}{\huvev} \lessapprox \frac{m_e}{m_t} \sim 10^{-5}
\end{equation}
For the rest of this section, we will assume that the vevs $\hdvev$ and $\huvev$ satisfy this condition and work in the large $\tan \beta$ limit of supersymmetric theories. The couplings in Eqn.~\eqref{InitialSuperPotential} can generate the flavor structure of the standard model only if they break all the flavor symmetries. The terms in  Eqn.~\eqref{InitialSuperPotential} break all flavor symmetries only if interactions involving the higgs triplet fields $\hutriplet$ and $\hdtriplet$ are present in the theory. For example, the only flavor breaking interactions of the first and second generation $U$ fields in  Eqn.~\eqref{InitialSuperPotential} arise through the couplings $\lambda_{1i} \hdtriplet U_1 D_i$ and $\lambda_{2i} \hdtriplet U_2 D_i$. These interactions must therefore play an essential role in generating the flavor hierarchies. Constraints from proton decay experiments \cite{Raby:2008pd} require these triplet fields to have masses $m_{\hutriplet}, \, m_{\hdtriplet} \gtrapprox \Mgut$.

 Quark and lepton masses arise from superpotential terms and are protected by SUSY non-renormalization theorems. Any radiative mechanism that generates these terms must therefore be proportional to the scale $\FSUSY$ of SUSY breaking. The necessity of the triplet higgs fields and SUSY breaking in this mechanism implies that the radiative masses generated will all be proportional to $\frac{\FSUSY}{m^{2}_{\hdtriplet}} \sim \frac{\FSUSY}{\Mgut^{2}}$. These radiative corrections can make significant contributions to the quark and lepton mass matrices only if $\FSUSY \sim \mathcal{O}\l\Mgut\r$. Consequently, we will assume that supersymmetry is broken at the GUT scale and that the weak scale is stabilized by fine tuning {\it i.e.} we work in the context of split SUSY \cite{ArkaniHamed:2004fb, ArkaniHamed:2004yi, Giudice:2004tc}. Since we work in the limit of large $\tan \beta$, we find ourselves in the parameter space of the theory where $h_u$ is fine tuned to be light with $\huvev \sim \text{TeV}$ and $h_d$ receives a GUT scale mass with $\hdvev \ll \huvev$.

In this context, the spurions $y_3$ and $\lambda$ breaks all flavor symmetries and hence, once the chiral symmetries forbidding fermion mases  are broken, it will generate masses for all the generations at some loop order.
After the first term in Eqn.~\eqref{InitialSuperPotential} is fixed to be in the $10_3$ direction, there is still a $U(2)_{10} \times U(3)_{\FiveBar}$ flavor symmetry.  This can be used to choose a basis in which the arbitrary matrix $\lambda$ has a `staircase' form 
\begin{equation}
\lambda = \left(\begin{matrix} \lambda_{11} & \lambda_{12} & 0 \\ 0 & \lambda_{22} & \lambda_{23} \\ 0 & 0 & \lambda_{33} \end{matrix}  \right)
\label{Eqn-staircase}
\end{equation}
This staircase form can be arrived at by using  $U(3)$ rotations on $\FiveBar_j$ to make $\lambda$ an upper triangular matrix, followed by $U(2)$ rotations on $10_i$ to get the zero on the upper right corner of the matrix.  In this basis, it is clear that the second generation masses will be generated from the third generation mass at a lower loop order than the first generation since the third generation can talk to the first generation only through the second generation.  

In the following, we first discuss the generation of the top yukawa coupling $y_3 10_3 10_3 H_u$ consistent with the flavor democratic principles outlined earlier in this paper. We then study the generation of the remaining quark and lepton masses, followed by estimates of wavefunction renormalizations. The subsequent sections address the effects of higher dimension Planck suppressed operators, GUT breaking operators and superpartner contributions. We then describe the parametric behavior of the final mass matrix obtained from this theory and address the issue of neutrino masses in this framework. We then conclude with a discussion on the requirements imposed on UV completions of this model into the framework of split SUSY.  The numerics of this model are discussed in Section \ref{PureSplitNumbers} where we also compute the color factors associated with the various loop diagrams that generate the fermion mass hierarchy. 

\subsubsection{The Top Yukawa}
\label{TopYukawa}
The challenge in generating the tree level yukawa term for the top quark in Eqn.~\eqref{InitialSuperPotential} is to generate this operator without corresponding terms for the first and second generation $10$'s but with a completely democratic treatment of all the flavors. We achieve this by initially forbidding the coupling of $10_i$ to the $H_u$ by a global $U(1)_H$ symmetry. $H_u$ will be allowed to couple to another sector of the theory to which the $10_i$ fields are linearly coupled ({\it i.e.} the vector $c_i$ in section \ref{DominoPhilosophy}) while preserving the $U(1)_H$ symmetry. Upon spontaneous breaking of the $U(1)_H$ symmetry forbidding the yukawa operators, these operators will be generated through the vector $c_i$. In the absence of other flavor breaking spurions in the $10_i \otimes 10_j$ space, the vector $c_i$ can be christened $10_3$. This mechanism will thus only yield the required yukawa coupling $y_3 \, 10_3 \, 10_3 \, H_u$.  Similar ideas have been employed previously (see e.g.~\cite{Dobrescu:2008sz, Froggatt:1978nt, Balakrishna:1987qd, Balakrishna:1988ks, Banks:1987iu}).

We demonstrate this idea with a concrete example involving the addition of one vector-like pair of fields in the 10 of SU(5), $\l 10_N, \TenBar_N\r$ and an SU(5) singlet $\sigma$ whose vev $\sigmavev$ will break the $U(1)_H$ symmetry forbidding the yukawa operators. We choose $U(1)_H$ charge assignments that do not allow the yukawa operators $10_i  \, 10_j \, H_u$  (see table  \ref{Tab:charges}) and write all the allowed $U(1)_H$ invariant superpotential terms
\begin{equation}
\mathcal{W} \supset \lambdac \, 10_i  \, \FiveBar_j \,  H_d \,+\, c_i\, \sigma \, 10_i \, \TenBar_N  \, + \,  M \, 10_N \, \TenBar_N \,+\, y_N \,10_N\, 10_N \, H_u
\label{UglyPotential}
\end{equation}
where $c_i$ is arbitrary vector and $\lambdac$ is an arbitrary tensor in flavor space.  A priori, we have the freedom to choose the charge $Q_d$ of $H_d$ independent of the charge of $H_u$. In the example shown in Table \ref{Tab:charges}, $Q_d$ was chosen to allow the operator $B_{\mu} \, H_u \, H_d$ (see equation \eqref{eqn: b mu term minimal model}) whose use will become apparent in section \ref{DownMasses}. However, this choice is not necessary since the operator $B_{\mu} \, H_u \, H_d$ can be generated after $U(1)_H$ is broken. 

We have not included the term $\mu H_u H_d$ for the higgs fields, even though it is allowed by the $U(1)_H$ charge assignments in Table  \ref{Tab:charges}. This term is forbidden by an $R$ symmetry that keeps the gauginos light in split SUSY. The low energy theory cannot contain the higgsino triplets since they spoil gauge coupling unification. These must be removed from the  spectrum without breaking the $U(1)_R$ symmetry. We discuss ways to achieve this goal in Section \ref{SplitTriplets}.

\begin{table}
\begin{center}
\begin{math}
\begin{array}{|c|c|c|}
\hline
\text{Field} & U(1)_H \text{ Charge}  & \text{Example} \\
\hline
\sigma &  +1 & +1 \\
\hline
10_i & Q_{10} & +1 \\
\hline
H_u & -2\l 1 + Q_{10} \r& -4 \\
\hline
H_d & 2 \l 1 + Q_{10}\r & +4 \\
\hline
\FiveBar_i & -\l 3  + 2\, Q_{10} \r  & -5 \\
\hline
10_N & -\l 1 + Q_{10}\r&-2\\
\hline
\end{array}
\end{math}
\caption[Example Charge Assignments]{\label{Tab:charges} $U(1)_H$ charge assignments that allow the terms in the superpotential Eqn.~\eqref{UglyPotential} and the $B_{\mu}$ term in Eqn.~\eqref{eqn: b mu term minimal model} but not the yukawa couplings $10_i \, 10_j \, H_u$. }
\end{center}
\end{table}

\begin{figure}
\begin{center}
\includegraphics[width=3 in]{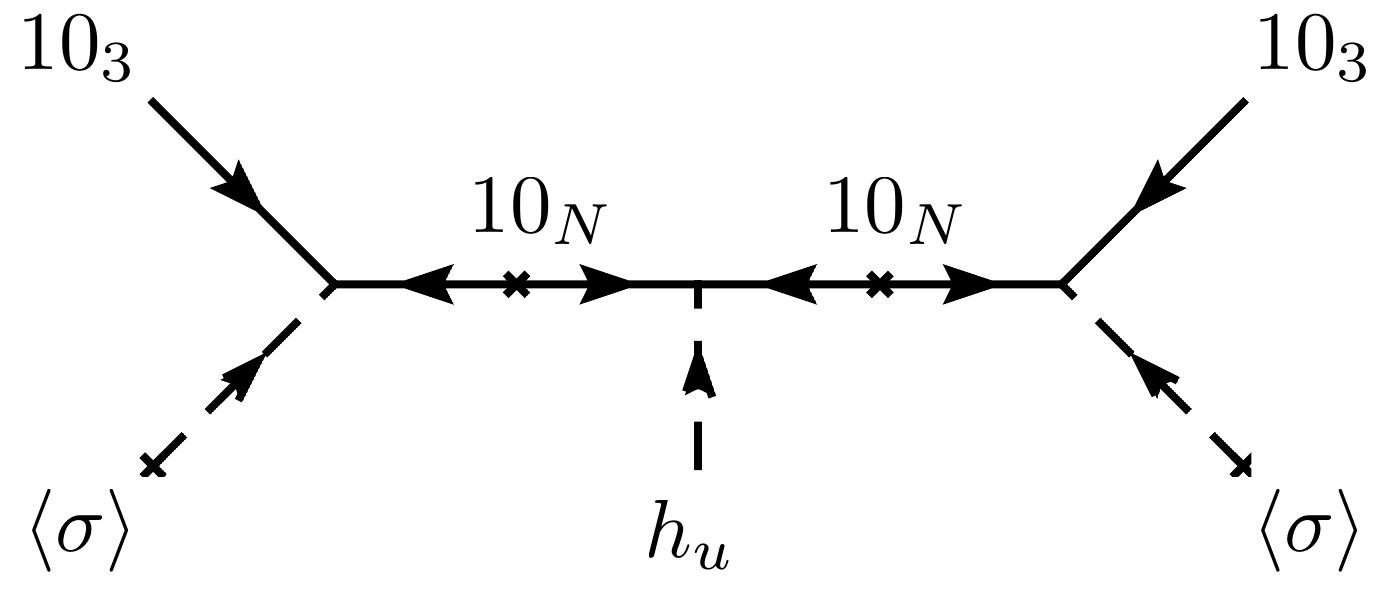}
\caption{ \label{fig-minfrog} The generation of the top yukawa coupling by integrating out the $10_N$ fields, using the operators $c_i \, 10_i \, \TenBar$, $M \, 10_N \, \TenBar_N$ and $y_N \, 10_N \, 10_N \, H_u$ in Eqn.~\eqref{UglyPotential}. }
\end{center}
\end{figure}

Integrating out the $10_H$ fields, the operator
\begin{equation} 
\frac{y_N^2 \sigma^2 \l c_i 10_i \r \l c_j 10_j \r H_u}{M^2} 
\end{equation}
 is generated  by the tree level diagram in figure (\ref{fig-minfrog}). The superpotential \eqref{UglyPotential} does not choose a direction other than $c_i$ in the $10_i \otimes 10_j$ flavor space. Thus we can choose a basis in this space in which the $10_3$ direction points along the direction chosen by the vector $c$. Using this basis to define flavor directions, we get the superpotential in Eqn.~\eqref{InitialSuperPotential} once $\sigma$ gets a vev with 
\begin{equation} 
y_3 =\l \frac{y_N \sigmavev }{M} \r^2
\end{equation}
This yukawa term is $\OrderOne$ when $y_N \sim \OrderOne$ and $\sigmavev \sim M$.

\subsubsection{Up Masses}
\label{UpMasses}
\begin{figure}[]
\begin{center}

\subfigure[ ]{\label{fig-2loopup}}
\includegraphics[width=3 in]{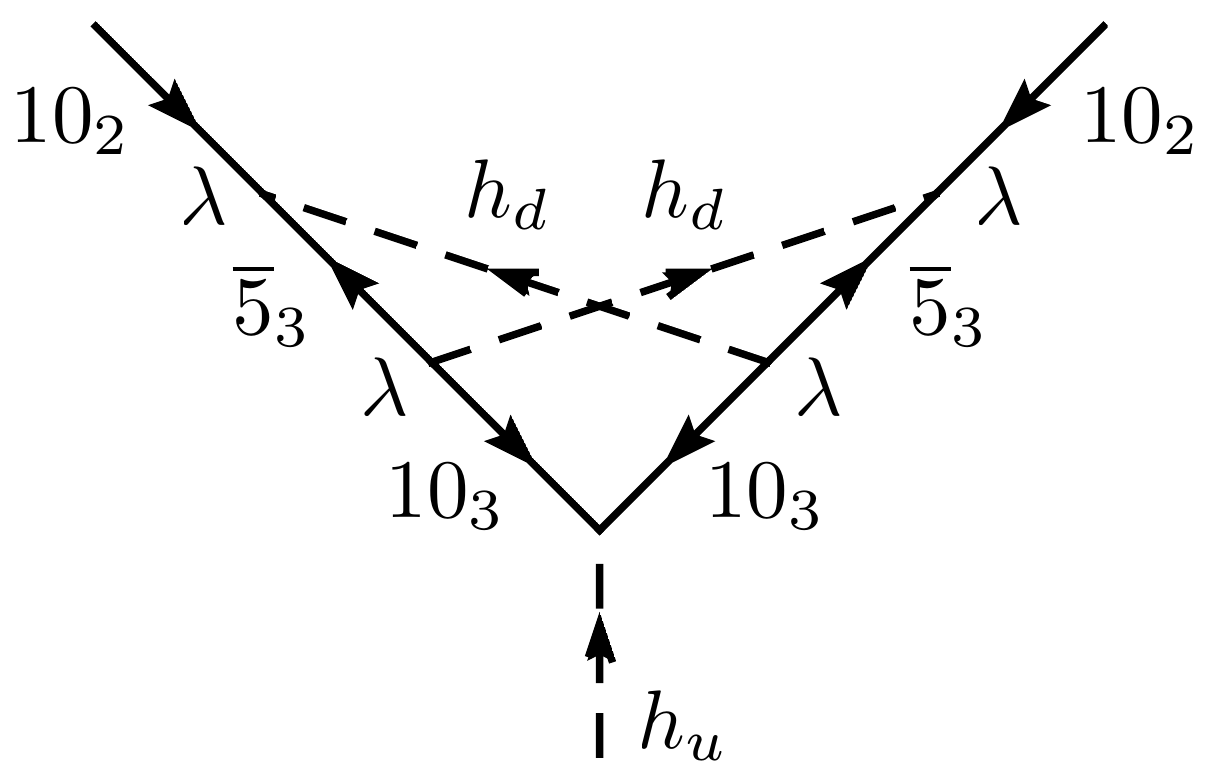} \qquad
\subfigure[ ]{\label{fig-2loopdown}}
\includegraphics[width=3 in]{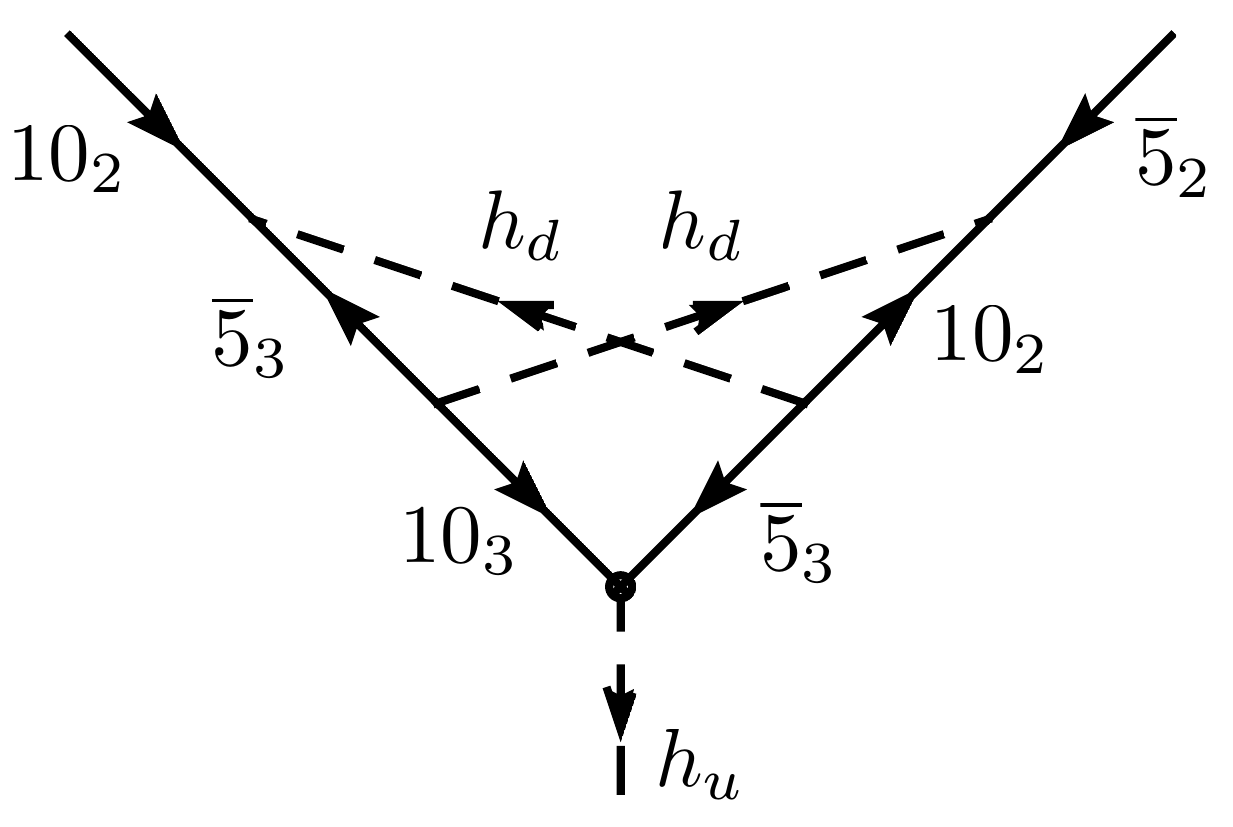}
\\
\subfigure[ ]{\label{fig-wavefunction10}}
\includegraphics[width=3 in]{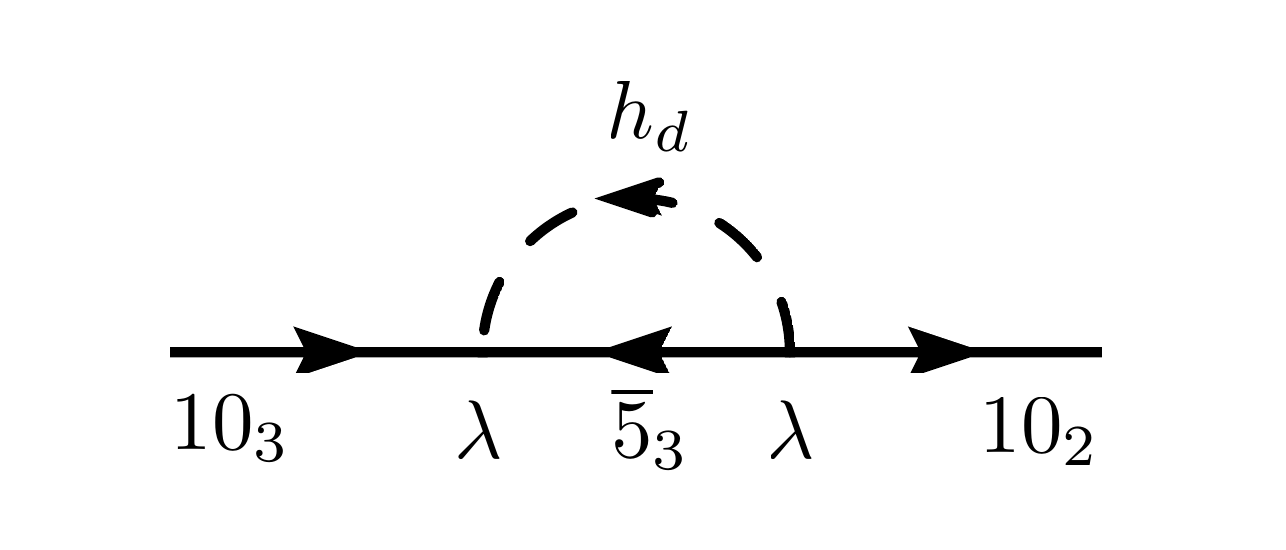}  \qquad \qquad
\subfigure[ ]{\label{fig-wavefunction5}}
\includegraphics[width=3 in]{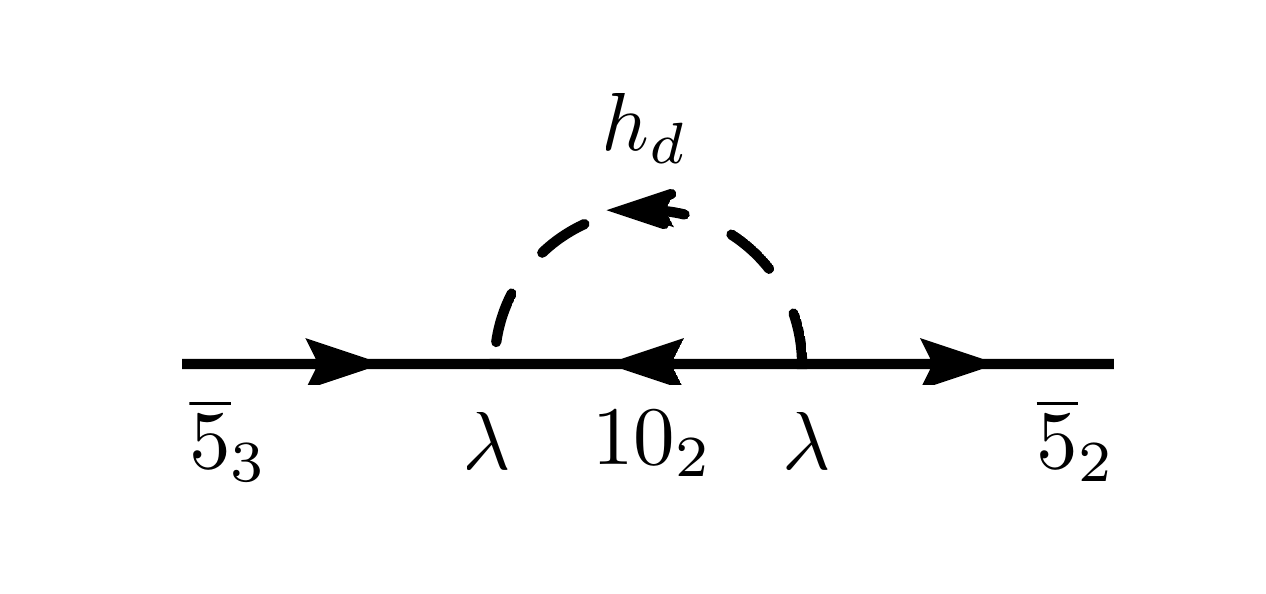}
\\
\subfigure[ ]{\label{fig-1loopbtau}}
\includegraphics[width = 2.0 in]{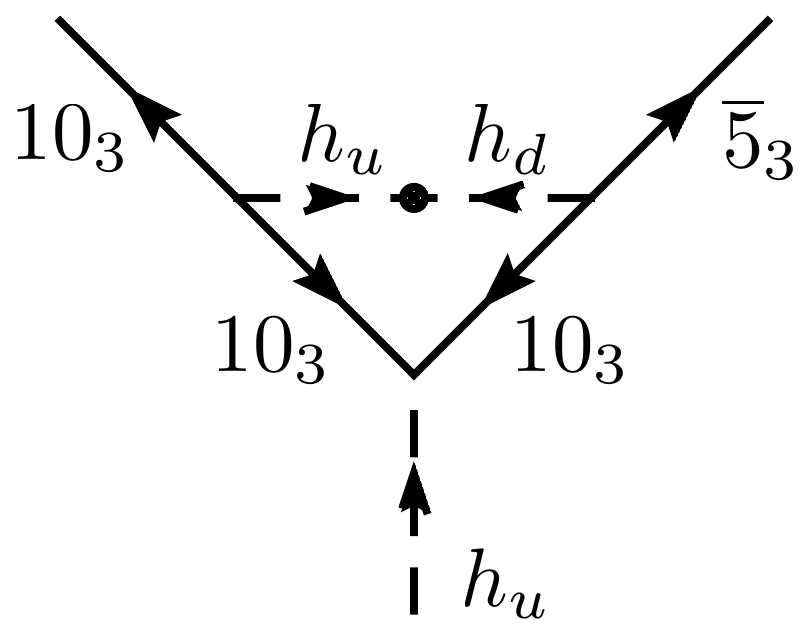}

\caption{ The core set of diagrams that generate the mass hierarchies in this model. Figures \ref{fig-2loopup} and \ref{fig-2loopdown} generate charm and $s - \mu$ masses at two loops below the top and $b-\tau$ masses respectively. Similar diagrams with the replacements $10_3 \rightarrow 10_2, \, 10_2 \rightarrow 10_1, \, \FiveBar_3 \rightarrow \FiveBar_2 \text{ and } \FiveBar_2 \rightarrow \FiveBar_1$,  contributes to the first generation masses at two loops below the second generation masses. In figures \ref{fig-wavefunction10} and  \ref{fig-wavefunction5} we show one loop kinetic mixings  $\TenDagger_2 \, 10_3$ and $\FiveBarDagger_2 \, \FiveBar_3$. Similar diagrams also generate kinetic mixing operators $\TenDagger_1 \, 10_2$, $\FiveBarDagger_1\, \FiveBar_2$ at one loop  with the replacements $10_3 \rightarrow 10_2$,  $10_2 \rightarrow 10_1$,  $\FiveBar_3 \rightarrow \FiveBar_2$ and  $\FiveBar_2 \rightarrow \FiveBar_1$.  In figure \ref{fig-1loopbtau}, we show  the generation of a $b-\tau$ mass at one loop below the top quark mass. The parametric hierarchy in the generated masses is as predicted by the spurion analysis of section \ref{Spurions}. }
\end{center}
\end{figure}

The top quark mass $y_3 \left(10_3 10_3 H_u\right)$ breaks the chiral symmetry forbidding fermion masses in the $10_i \otimes 10_j$ space ({\it i.e.} the up quark sector). In the presence of this term, the spurion $\lambda$, which breaks all the other remaining flavor symmetries, will generate masses for the first and second generation up quarks. Starting with the top mass, the diagram  in Figure \ref{fig-2loopup} generates a second generation charm mass $y_2 \left(10_2 10_2 H_u\right)$ at 2 loops. This diagram is a non-planar, log divergent diagram whose magnitude is (see Section \ref{PureSplitNumbers})
\begin{equation}
y_2 \sim y_3 \left(\lambdac\right)^4  N_2^u \left(\loopfactor\right)^2 \log\left(\frac{\Lambda^2}{m_{h_d}^2}\right)
\label{Eqn-charm}
\end{equation}
where $\lambdac$ are elements of the $\lambda$ matrix,  $N_2^u$ is the color factor for this diagram and $\Lambda$ is the UV cut-off of the diagram.  Of course, in this particular case of generating the $10_2 10_2$ element of the Yukawa matrix, in place of $\lambdac^4$ in Eqn.~\eqref{Eqn-charm} we really mean $\lambda^2_{33} \lambda^2_{23}$.  In the limit of unbroken supersymmetry, the diagram in  Figure \ref{fig-2loopup} is cancelled by its counterparts involving super-partners. When supersymmetry is broken, the diagrams will be cut off at the SUSY breaking scale $\FSUSY$. The diagram also falls apart when the momentum through the loops are larger than the scale $M$ at which the effective top yukawa $y_3 \left(10_3 10_3 H_u\right)$  is generated (see subsection \ref{TopYukawa}). This loop is therefore cut off at the scale $\Lambda$ which is the smaller of these two scales. For the purposes of this paper, we will assume that $\Lambda \sim \FSUSY \sim \Mgut$, a condition that is satisfied as long as $M \gtrapprox \Mgut$. Numerically, the yukawas in Eqn.~\eqref{Eqn-charm} generate the observed charm mass, $\frac{y_3}{y_2} \approx 300$, for $\lambdac \approx 1.5$ when the log in (\ref{Eqn-charm}) is $\approx 4$ and the color factor $N^u_2 \approx 4$. Note that this comparison is made at the scale $\Lambda \sim \Mgut$ at which these yukawas are generated. 

\begin{figure}
\begin{center}
\includegraphics[width=3 in]{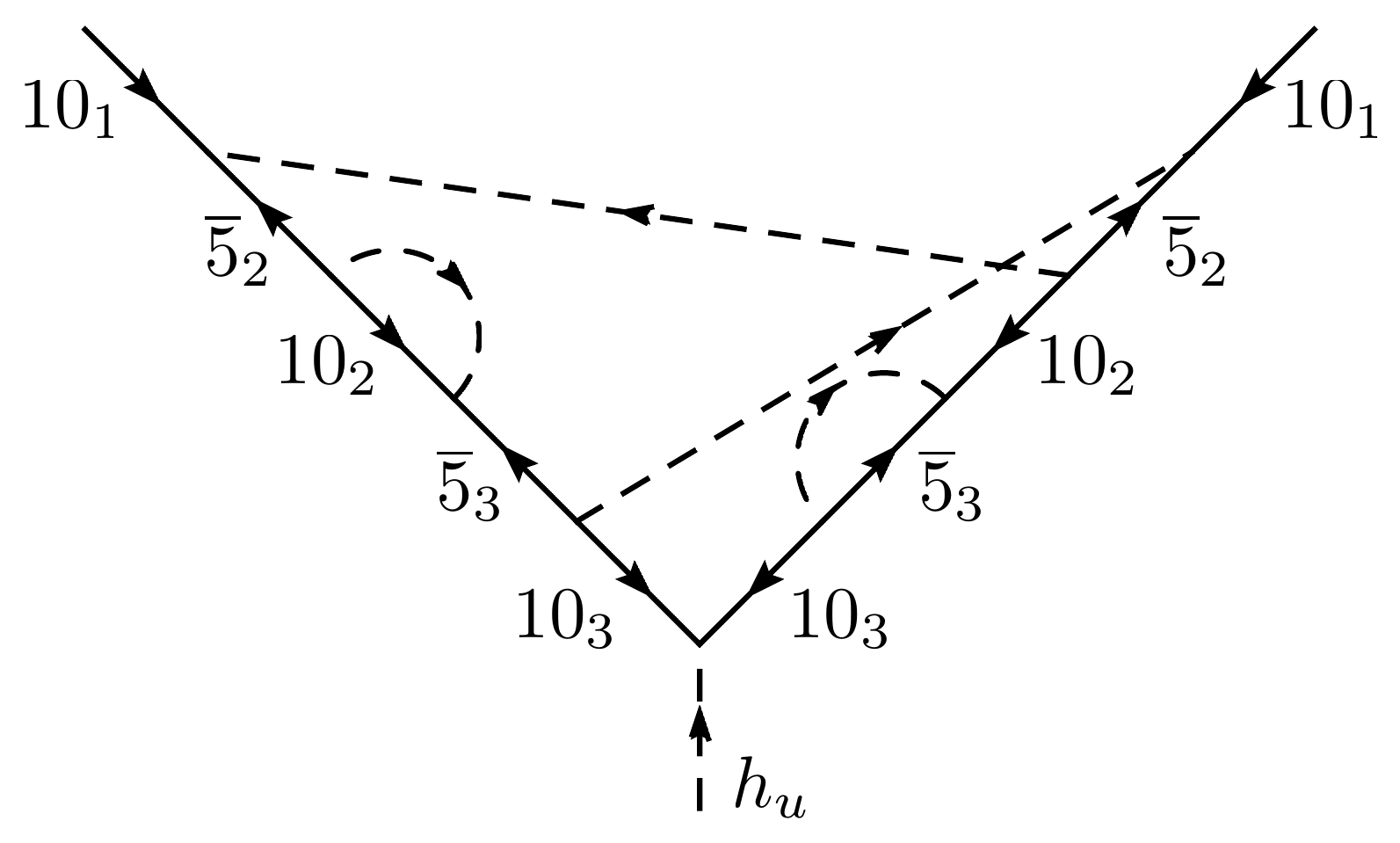}
\caption{ \label{fig-4loopup} A four loop contribution to the up quark mass $10_1 \, 10_1 \, H_u$. There are four more such diagrams that contribute to the up quark mass in addition to the effective four loop contribution that arises from figure \ref{fig-2loopup}.}
\end{center}
\end{figure}
As emphasized earlier, the staircase structure of the $\lambda$ matrix ensures that first generation masses are not generated at 2 loops from the top mass by this diagram. However, once the charm mass is generated, the diagram in Figure \ref{fig-2loopup} will generate an up quark mass   $y_1 \left(10_1 10_1 H_u\right)$ at 2 loops from the charm mass. In addition to these effective four loop diagrams, there are 5 more diagrams like Figure \ref{fig-4loopup} that also contribute to the up  mass. We expect these diagrams to contribute to the up mass at comparable levels and estimate the total contribution to be 
\begin{equation}
y_1 \sim 6 \left( y_c \left(\lambdac\right)^4 N_2^u \left(\loopfactor\right)^2 \log\left(\frac{\Lambda^2}{m_{h_d}^2}\right)\right)
\label{Eqn-up}
\end{equation}
This estimate is arrived at by multiplying the total number of 4 loop contributions to the up mass ({\it i.e.} 6) and the value of the two loop contribution to the up mass from the charm mass (obtained from equation \eqref{Eqn-charm}). This contribution is comparable to the up mass when $\lambdac \sim 1$. 

Borrowing the language of section \ref{Spurions}, we see that the domino mechanism has generated a charm mass at order $\epsilon^2$ ({\it i.e} 2 loops) and an up mass at order $\epsilon^4$ ({\it i.e.} 4 loops), as expected from the spurion analysis. In addition to the masses, the diagrams discussed in this section also contribute to the $23$ mixing angle at order $\epsilon^2$ and the $12$, $13$ mixing angles at order $\epsilon^4$. However, as expected from the spurion analysis, these are sub-dominant to the lower order contributions to the mixing angles from wavefunction renormalizations (see section \ref{Zs}).  

\subsubsection{Down and Lepton Masses}
\label{DownMasses}
The chiral symmetry forbidding fermion masses in the $10_i \otimes \FiveBar_j$ space is broken by the operators $10_i \FiveBar_j h_d$ and $10_i \FiveBar_j \huDagger$. Since  $\hdvev \ll \huvev$ in this model, masses for the down quarks and leptons must be generated through the operators $10_i \FiveBar_j \huDagger$. This operator is not invariant under SUSY and can only be generated after SUSY breaking. The simplest SUSY breaking operator that connects the $10_i \FiveBar_j H_d$ operators in the superpotential (\ref{InitialSuperPotential}) to the chiral symmetry breaking term $y_3 \left(10_3 10_3 H_u\right)$ is
\begin{equation}
\mathcal{L}_{\text{\sout{SUSY}}} \supset B\mu \, H_u H_d
\label{eqn: b mu term minimal model}
\end{equation}

\begin{figure}
\begin{center}
\includegraphics[width=4 in]{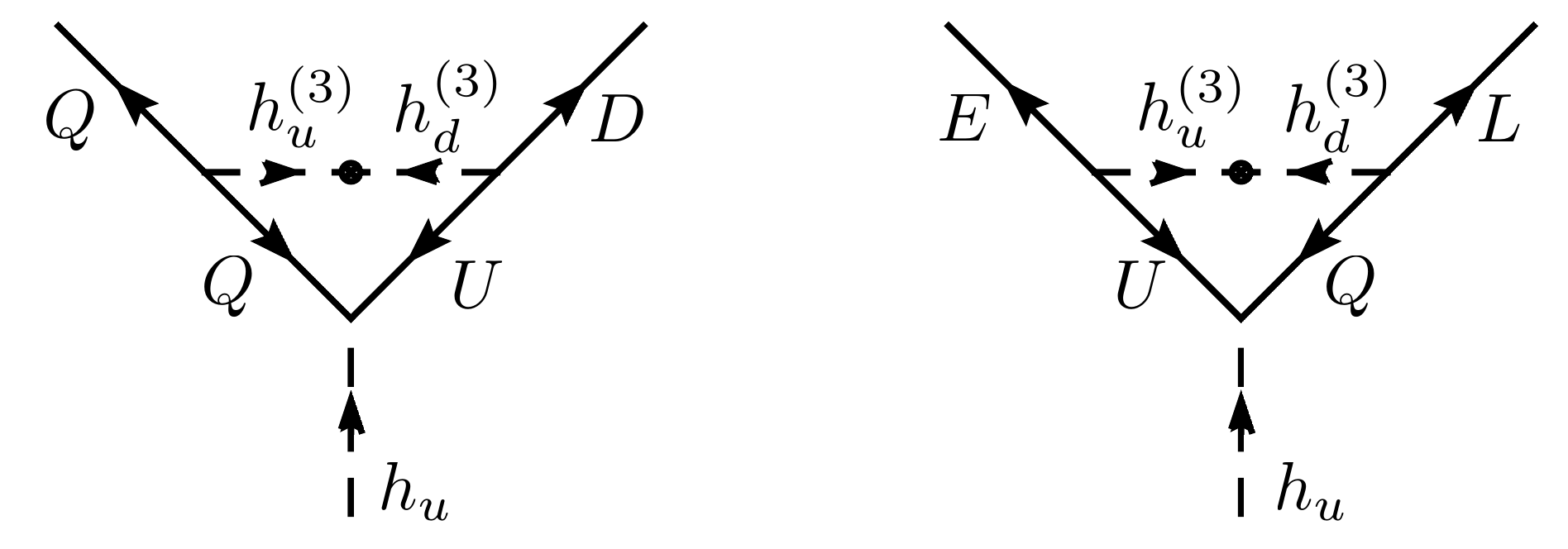}
\caption{ \label{fig-tripletbtau} Generation of $SU(5)$ breaking masses for the $b$ and $\tau$ at one loop below the top mass using a $SU(5)$ breaking $B_{\mu}$ term $\tripletbmu \, \hutriplet \, \hdtriplet$ for the triplets.}
\end{center}
\end{figure}

In the presence of this operator, a mass $x_3 \left(10_3 \FiveBar_3 \huDagger\right)$ is generated for the $b$ quark and $\tau$ lepton at one loop from the top mass through the diagram in figure (\ref{fig-1loopbtau}). The value of this diagram when $B\mu \lessapprox m_{h_d}^2$ is 
\begin{equation}
x_3 \sim y_3 \left(y_3 \lambdac\right) N^d_1 \left(\loopfactor\right) \frac{B\mu}{m_{h_d}^2}
\label{Eqn:btau}
\end{equation}
where  $N^d_1$ is the color factor for the diagram. In the absence of $SU(5)$ breaking,  this diagram will give equal masses to both the $b$ and $\tau$. Under the standard model gauge interactions, the $B\mu$ term decomposes into 
\begin{equation}
B_{\mu} H_u H_d \supset \doubletbmu h_u h_d + \tripletbmu \hutriplet \hdtriplet
\end{equation}
where $h_u$, $h_d$ are the higgs doublets, $\hutriplet$, $\hdtriplet$ are higgs triplets and $\doubletbmu$, $\tripletbmu$ are their respective $B_{\mu}$ couplings. When $h_u$ develops a vev $\huvev$, the linear coupling $\doubletbmu h_u h_d$ causes $h_d$ to develop a vev $\hdvev \sim \frac{\doubletbmu}{m_{h_{d}}^2} \huvev$. Since our scenario requires $\frac{\hdvev}{\huvev} \lessapprox 10^{-5}$, we must have $\frac{\doubletbmu}{m_{h_{d}}^2} \lessapprox 10^{-5}$. Consequently, the contributions of the higgs doublets to the $b$ and $\tau$ masses in Figure \ref{fig-1loopbtau} are suppressed. In the limit of unbroken $SU(5)$, we will have $\doubletbmu \sim \tripletbmu$ and hence contributions to Figure \ref{fig-1loopbtau} from the higgs triplets will also be suppressed resulting in unacceptably small masses for the $b$ and $\tau$. However, since $SU(5)$ is broken at this scale, it is possible for the $B_{\mu}$ term to violate $SU(5)$ invariance,  resulting in $\doubletbmu \ll \tripletbmu \sim m_{h_{d}}^2$. This splitting is reminiscent of doublet-triplet splitting in $SU(5)$ and it is conceivable that the physics responsible for this splitting could also split $\doubletbmu$ and $\tripletbmu$.  Once this has been achieved, loops involving $\tripletbmu$ will not be suppressed and the diagrams in figure (\ref{fig-tripletbtau}) will generate masses for $b$ and $\tau$ at one loop below the top quark mass. As a result of the explicit $SU(5)$ breaking, the $b$ and $\tau$ masses  generated by this mechanism will not be equal. These $SU(5)$ breaking effects are further studied in sections \ref{GUTBreaking} and \ref{PureSplitNumbers}. With $\frac{\tripletbmu}{m^{2}_{h_d}} \sim \OrderOne$ and a color factor $N^{d}_1 \sim 3$, the contribution from Eqn.~\eqref{Eqn:btau} is comparable to the $b-\tau$ mass for $\lambdac \sim 0.7$.

\begin{figure}
\begin{center}
\includegraphics[width=2 in]{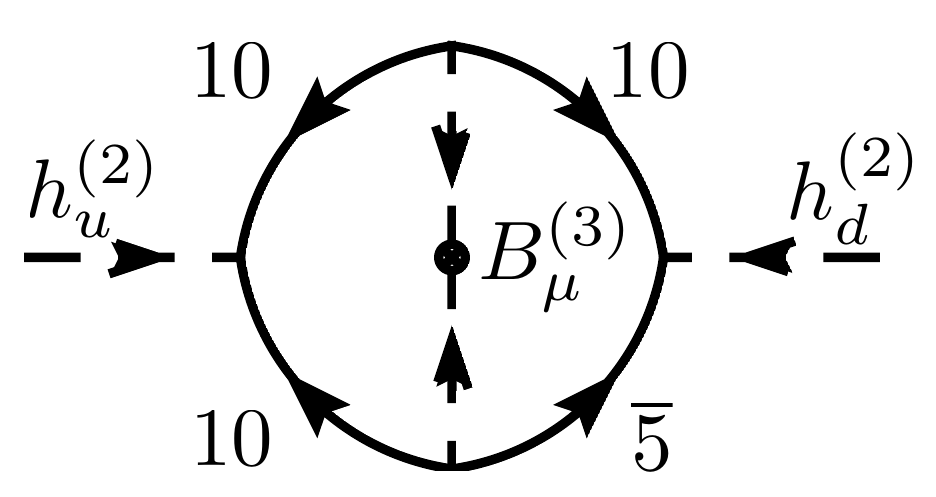}
\caption{ \label{fig-huhdmixing} A two loop contribution to the doublet $B_{\mu}$ term $\doubletbmu \, h_u \, h_d$.}
\end{center}
\end{figure}

We note that the generation of the $b - \tau$ yukawa coupling $10_3 \, \FiveBar_3 \, \huDagger$ will result in a contribution to  $\doubletbmu \sim \l \loopfactor \r \, x_3 \, \lambda_{33} \, \Lambda^2$ from the two loop diagram of figure \ref{fig-huhdmixing}. This effective two loop contribution naturally leads to $\hdvev \sim \l \mathcal{O}\l 10^{-4} \r - \mathcal{O}\l 10^{-3} \r \r \huvev$ which is larger than the required $\hdvev \lessapprox \mathcal{O} \l 10^{-5} \r \huvev$. Consequently, this contribution must be tuned away during the initial tuning required to get $\doubletbmu \lessapprox \mathcal{O} \l 10^{-5} \r m^{2}_{h_d}$.

\begin{figure}
\begin{center}
\subfigure[]{\label{fig-3loop}}
\includegraphics[width=3 in]{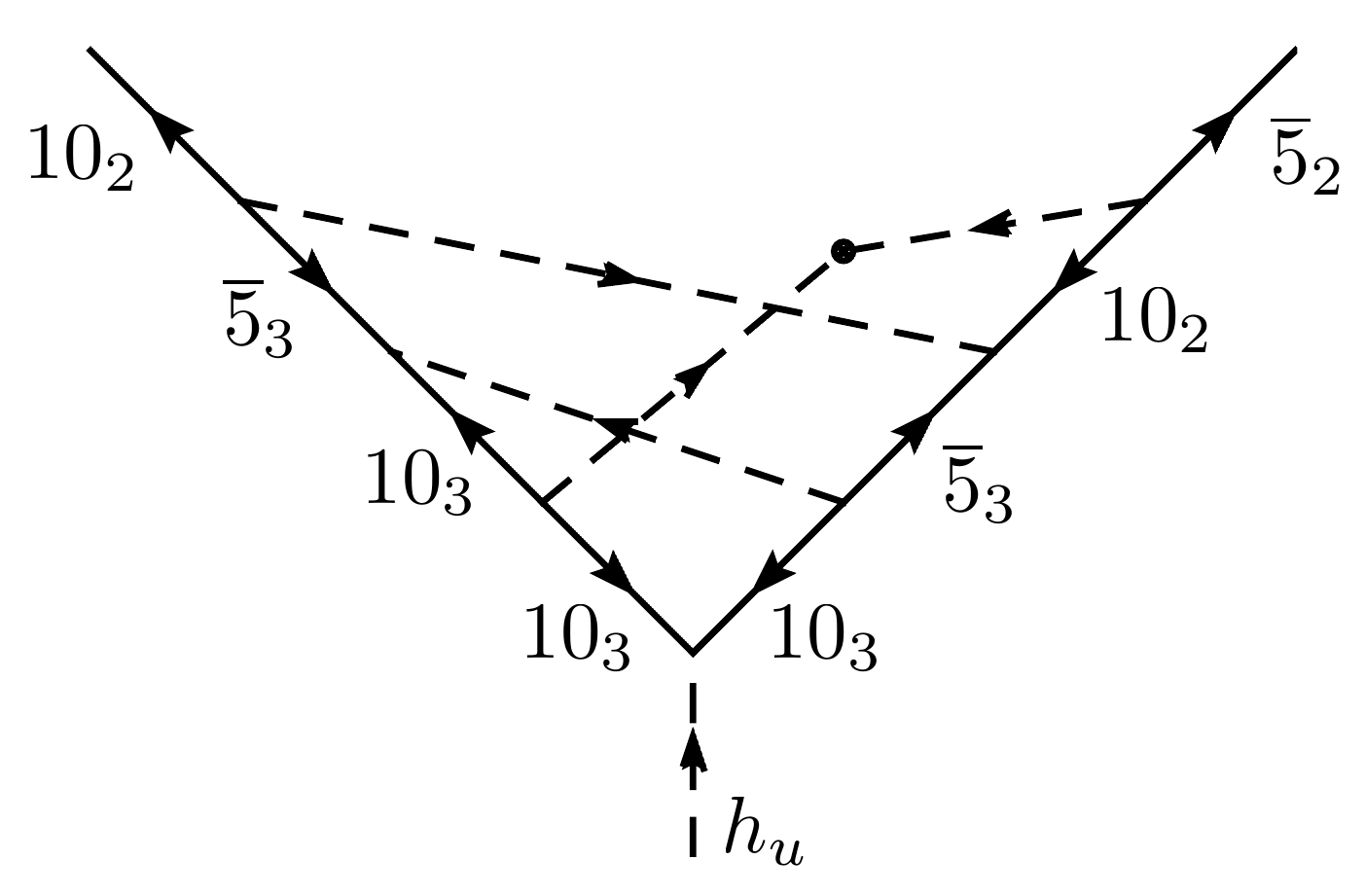} \qquad \qquad
\subfigure[]{\label{fig-2loopnew}}
\includegraphics[width=2.4 in]{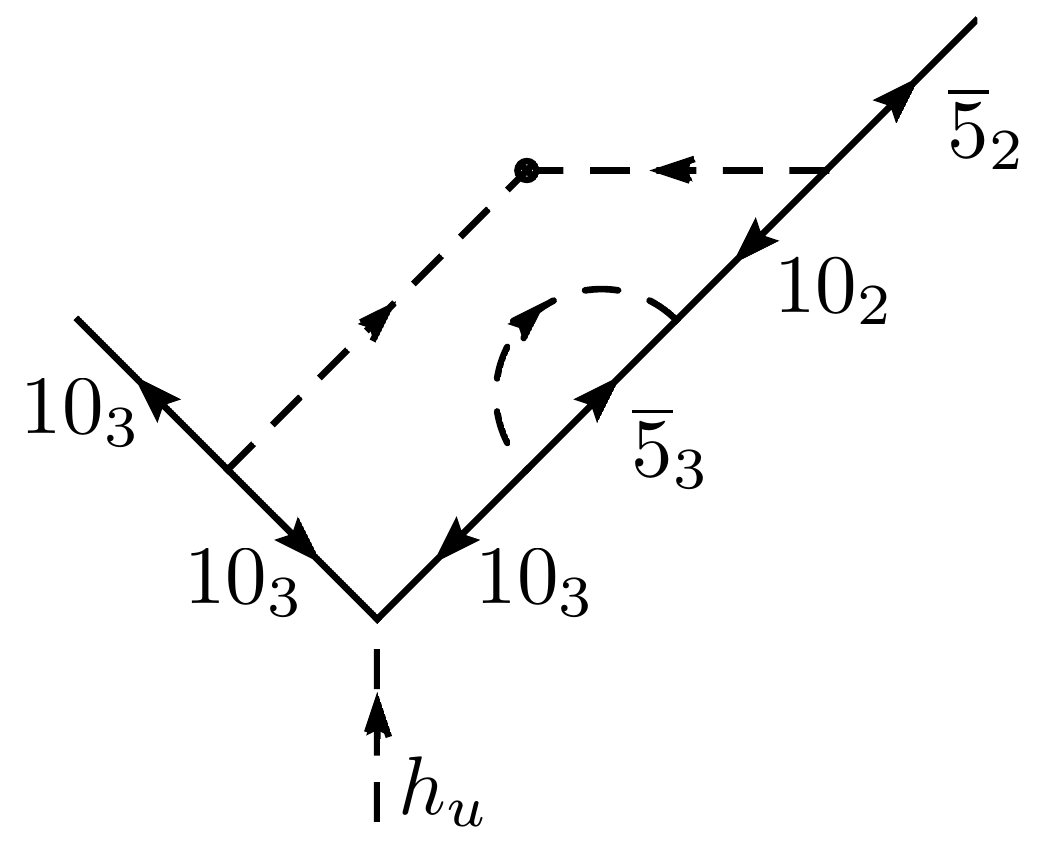}
\caption{ Fig. \ref{fig-3loop} is a contribution to the $s -\mu$ mass $10_2 \, \FiveBar_2 \, \huDagger$ at three loops below the top quark mass.  Fig. \ref{fig-2loopnew} gives a mixing $10_3 \, \FiveBar_2 \, \huDagger$ between the third and second generation down quark and leptons at two loops below the top mass. This mixing is also predicted by the spurion analysis of section \ref{Spurions}.}
\end{center}
\end{figure}

The $b$ and $\tau$ masses and the spurion $\lambdac$ break the chiral and flavor symmetries protecting the masses of the remaining down quarks and leptons. These masses will now be generated at some loop order. In particular, the $s-\mu$ masses, $x_2 \l10_2 \FiveBar_2 \huDagger\r$, are generated at two loops below the $b-\tau$ mass by the diagram in figure \ref{fig-2loopdown}.   This non-planar diagram is similar to the 2 loop contribution to the charm mass computed in equation \eqref{Eqn-charm} and its value can be estimated to be 
\begin{equation}
x_{2}  \sim x_{3} \left(\lambdac\right)^4  N_2^d \left(\loopfactor\right)^2 \log\left(\frac{\Lambda^2}{m_{h_d}^2}\right)
\label{Eqn-muon1}
\end{equation}
where $N_2^d$ is the color factor for this diagram. In addition to the above effective three loop contribution through the $b-\tau$ mass, there is also a direct three loop contribution from the top mass to $x_2$ from the diagram in figure \ref{fig-3loop}. This diagram involves the SUSY breaking $B\mu$ term and is estimated to be 
\begin{equation}
x_{2}  \sim y_{3} \left(y_{3} \lambda^2\right)  \left(\lambda\right)^3  N_3^d \left(\loopfactor\right)^3 \log\left(\frac{\Lambda^2}{m_{h_d}^2}\right) \frac{\tripletbmu}{m_{h_d}^2}
\label{Eqn-muon2}
\end{equation}

The color factors for these diagrams are $\sim 5$ (see section \ref{PureSplitNumbers}) and these contributions produce the correct second generation masses for $\lambdac \sim 2.5$. The diagram of figure \ref{fig-2loopnew} produces a mixing $x_{23}$ between the second and third generation at two loops. This planar diagram also involves the SUSY breaking $B\mu$ term and yields 

\begin{equation}
x_{23} \sim y_3 \left(y_3\right) \left(\lambdac\right)^{3}\left(\loopfactor\right)^2\log\left(\frac{\Lambda^2}{m_{h_d}^2}\right) \frac{\tripletbmu}{m_{h_d}^2}
\label{mixing23}
\end{equation}

\begin{figure}
\begin{center}
\includegraphics[width=3 in]{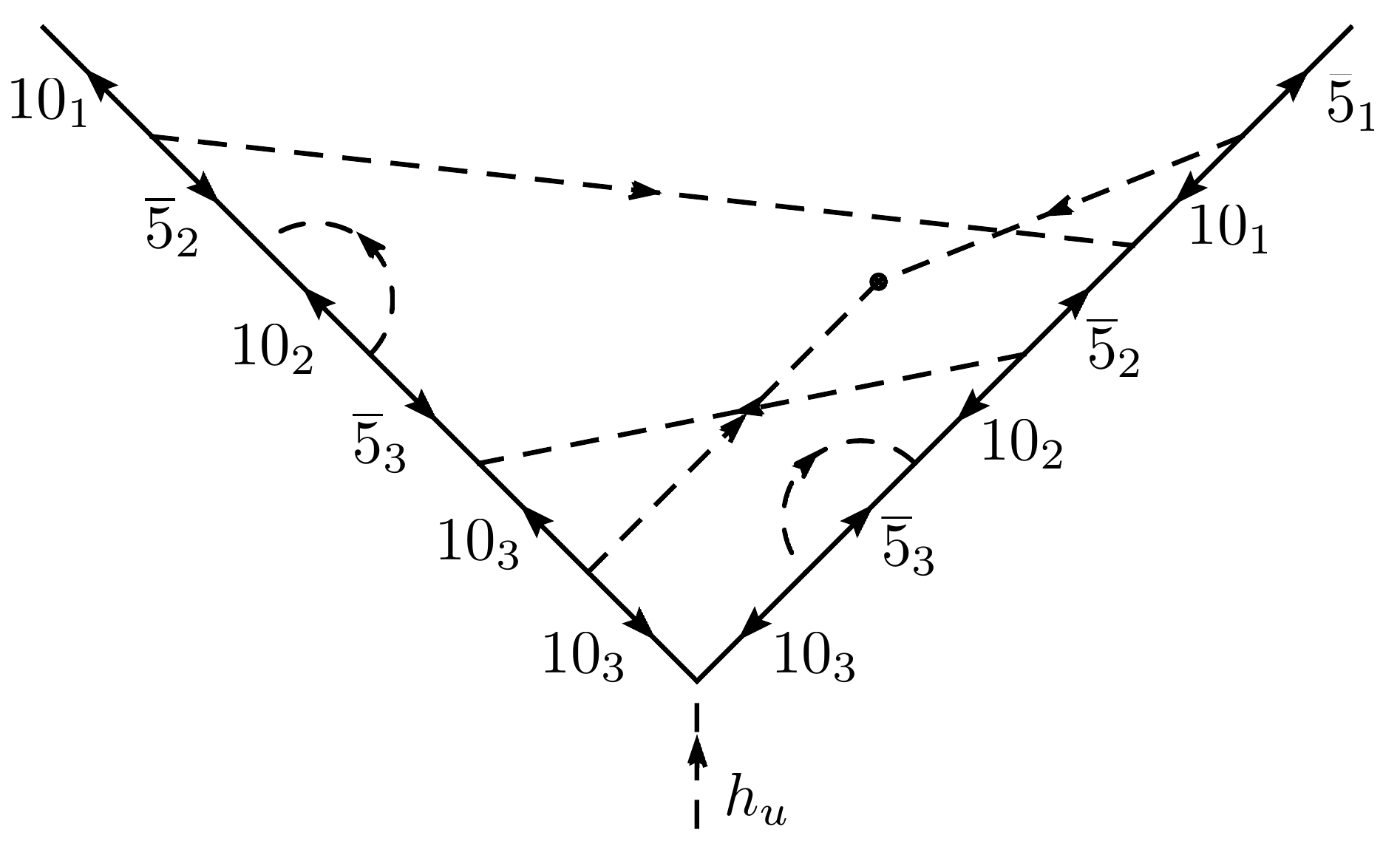}
\caption{\label{fig-5loopdown} A five loop contribution to the $d - e$  mass $10_1 \, \FiveBar_1 \, \huDagger$. There are ten more such diagrams that contribute to the $d-e$  mass in addition to the effective five loop contribution that arises from figure \ref{fig-2loopdown}. }
\end{center}
\end{figure}

The staircase form of the $\lambdac$ matrix (see equation ($\ref{Eqn-staircase}$)) ensures that the first generation does not receive a mass at this loop order. A first generation mass $x_1\l 10_1 \FiveBar_1 \huDagger\r$ will however be generated at two loops below the second generation mass by the diagram in figure \ref{fig-2loopdown}. In addition to these effective five loop diagrams, there are 11 more  diagrams like figure (\ref{fig-5loopdown}) that also contribute to the $d-e$ mass. We expect these contributions to be comparable and estimate the total contribution to be 
\begin{equation}
x_{1} \sim 12 \left(x_{2} \left(\lambdac\right)^4 N^{d}_2 \left(\loopfactor\right)^2 \log\left(\frac{\Lambda^2}{m_{h_d}^2}\right) \right)
\end{equation}

This estimate is arrived at by multiplying the total number of 5 loop contributions to  $x_1$ ({\it i.e.} 12) and the value of the two loop contribution to this mass from the second generation mass (obtained from equation (\ref{Eqn-muon1})). The $d-e$ mass  is generated from this mechanism with $\lambdac \sim 1$. 

The domino mechanism has generated a $b-\tau$ mass at order $\delta$ {\it i.e} at one loop, using the $\tripletbmu$ that communicates the chiral symmetry broken in the $10_i \otimes 10_j$ space to the $10_i \otimes \FiveBar_j$ space.  $s-\mu$  and $d-e$ masses are then generated at orders $\delta \epsilon^2$ and $\delta \epsilon^4$ respectively.  This hierarchy in the generated masses is in line with the expectations of the spurion analysis of section \ref{Spurions}. Other than the $23$ mixing generated by Eqn.~\eqref{mixing23} at order $\delta \epsilon$, the other diagrams discussed in this section contribute to mixing angles at higher loop orders than wavefunction renormalization (see section \ref{Zs}) and are hence sub-dominant. 

\subsubsection{Wave-function Renormalization}
\label{Zs}
Wave-function renormalizations cannot change the rank of the mass matrix and hence do not generate mass terms. However, once the masses have been generated, they can alter the magnitudes of the mass eigenvalues. They can also contribute to mixing angles between generations. The one loop diagrams in Figures \ref{fig-wavefunction10} and \ref{fig-wavefunction5} generate corrections to the operators $\FiveBarDagger_i \FiveBar_i$ and $\TenDagger_i 10_i$ in addition to the Kahler mixing operators  $\FiveBarDagger_1\FiveBar_2, \, \FiveBarDagger_2\FiveBar_3, \, \TenDagger_1 10_2, \text{ and } \TenDagger_2 10_3$ with magnitude 
\begin{equation}
\l\loopfactor\r N^{\l10, \FiveBar\r}_Z \l\lambdac\r^2 \log \l \frac{\Lambda^2}{m^2_{h_d}}\r
\label{EquationZ}
\end{equation}
where $N^{\l10, \FiveBar\r}_Z$ is the color factor for the diagram, whose value depends upon the propagating external state ({\it i.e.} $10 \text{ or } \FiveBar$). Kinetic mixing between the third and first generation ({\it i.e.} the operators $\TenDagger_1 10_3$ and $\FiveBarDagger_1 \FiveBar_3$) are induced at two loops by a similar mechanism.

\subsubsection{Planck Slop}
\label{PlanckSlop}
The $U(1)_H$ symmetry imposed earlier (in section \ref{TopYukawa}) to forbid the tree level yukawa couplings $H_u 10_i 10_j$ allows the higher dimension operator

\begin{equation}
\int d^2 \theta \frac{\sigma^2 H_u 10_i 10_j}{M^2_{pl}}
\label{EquationUpPlanckSlop}
\end{equation}
It is natural to suppress this dimension 6 operator by the cut-off of the theory {\it{i.e.}} the Planck scale $M_{pl}$. When $\sigma$ acquires a vev $\sigmavev$, this operator will make contributions $\sim \l \frac{\sigmavev}{M_{pl}}\r^2 \huvev$ to the up sector quark masses. For $\sigmavev \sim  \Mgut \approx 10^{16} \text{ GeV}$, this term is comparable to the mass of the first generation $u$ quark. 

The only flavor directions in the renormalizable part of the superpotential (\ref{InitialSuperPotential}) are the vector $c_i$ that picks the  $10_3$ direction and the tensor $\lambda$. The dimension six operator discussed above is the first source of new flavor directions in the theory consistent with the $U(1)_H$ symmetry imposed on the superpotential (\ref{UglyPotential}). As discussed above, its effects are comparable to the masses of the first generation. However, if $U(1)_H$ were a global symmetry, one may worry about the generation of  $U(1)_H$ violating lower dimensional operators such as $10_i 10_j H_u$ and $\sigma 10_i 10_j H_u$ that would introduce other new flavor directions in the theory. These operators could be generated by Planckian physics since global symmetries are believed to be violated by quantum gravity. However, we are not perturbed by this possibility since $U(1)_H$ could in fact be gauged in this model without affecting the domino mechanism. It is also straightforward to find discrete symmetries that allow the operators in (\ref{UglyPotential}) while forbidding the dangerous lower dimension operators discussed in this paragraph.  The global $U(1)_H$ symmetry then emerges as an accidental consequence of these discrete symmetries. 

Due to the existence of the diagram Figure \ref{fig-minfrog}, every charge assignment consistent with Eqn.~\eqref{UglyPotential} will allow the operator in Eqn.~\eqref{EquationUpPlanckSlop} and hence this operator is likely to be generated by any UV completion of this theory. Thus ``planck slop" in the up sector masses is unavoidable. It is also possible to find charge assignments ({\it e.g} $Q_{10} = 0$ in Table \ref{Tab:charges}) that allow the dimension six operator 

\begin{equation}
\int d^4 \theta \frac{\sigma^{\dagger} \HuDagger 10_i \FiveBar_j}{M^2_{pl}}
\label{EquationDownPlanckSlop}
\end{equation}
without allowing the dangerous operators discussed above. This operator contributes to down and lepton masses if $\sigma$ acquires a SUSY breaking $F$ term vev $\sim \langle F_{\sigma} \rangle$. This contribution is also of order the electron and down quark masses for $ \langle F_{\sigma} \rangle \sim \l \Mgut \r^2$. These arguments suggest that the masses and mixing angles of the first generation quarks and leptons are susceptible to the effects of additional flavor directions generated by Planckian physics.  Interestingly, such Planckian physics naturally generates masses, mixings, and the phase of the CKM matrix at their observed levels, as we will see in Section \ref{Sec: numerical results}.

\subsubsection{GUT Breaking}
\label{GUTBreaking}
$SU(5)$ breaking effects in the higgs sector, such as a mass splitting between the doublet and triplet components of $H_d$, feed into the down quark and lepton masses generated by the domino mechanism. The model discussed in this section requires $\doubletbmu \ll \tripletbmu$ (see section \ref{DownMasses}). In this case, the dominant source for the $b$ and $\tau$ masses are the triplet components of $H_d$.  The color factor $N^b_1$ for the diagram in figure \ref{fig-tripletbtau}  that generates the $b$ mass is different from the color favor $N^{\tau}_1$ of the diagram responsible for the $\tau$ mass. Following equation (\ref{Eqn:btau}), the ratio 
\begin{equation}
\frac{m_{\tau}}{m_b}  = \frac{N^{\tau}_1}{N^{b}_{1}} = \frac{3}{2}
\label{brokenbtau}
\end{equation}
Note that this depends only on the ratio of the color factors and not on the values of couplings or anything else in the diagram.  This is close to the ratio  $\frac{y_\tau}{y_b} \approx 1.7$ expected in split SUSY \cite{Giudice:2004tc}.  Masses for the other generations are generated by the diagrams \ref{fig-2loopdown}, \ref{fig-3loop},  \ref{fig-2loopnew} and \ref{fig-5loopdown}. $SU(5)$ breaking enters these diagrams through mass splittings between the doublet and triplet components of $H_d$. The color factors and mass ratios that control the sizes (for example, see equation (\ref{Eqn-muon1})) of these diagrams are determined by the component of the $H_d$ that runs through these diagrams. $\OrderOne$ doublet-triplet splitting will lead to $\OrderOne$ differences in the generated masses. In section \ref{PureSplitNumbers}, we will show that these higgs sector doublet-triplet splittings  can naturally accommodate the observed violations of GUT relations in the down and lepton masses.

\subsubsection{Superpartner Effects}
\label{superpartner}
In the limit of unbroken SUSY, the radiative corrections that generate the fermion masses are cancelled by corresponding superpartner loops. SUSY breaking is thus an essential part of this mechanism. But, these SUSY breaking terms ({\it e.g.} scalar masses and $A$ terms) can potentially introduce new flavor spurions into the model. These new spurions cannot affect the hierarchy between the third and the second generation since the chiral symmetry breaking directly experienced by the top quark has to be communicated to the other generations. In this model, the breaking of the chiral symmetry is communicated through loops and hence the masses of the second generation will always be parametrically smaller than that of the third generation. However, as articulated in section   \ref{Spurions},  the first generation masses are parametrically smaller than the second generation masses  only because there are exactly two dominant flavor breaking spurions ({\it i.e.} $c$ and $\lambda$) in (\ref{UglyPotential}). The inclusion of new flavorful spurions through SUSY breaking terms can potentially upset the hierarchy between the first and second generations. 

\begin{figure}
\begin{center}
\includegraphics[width=3 in]{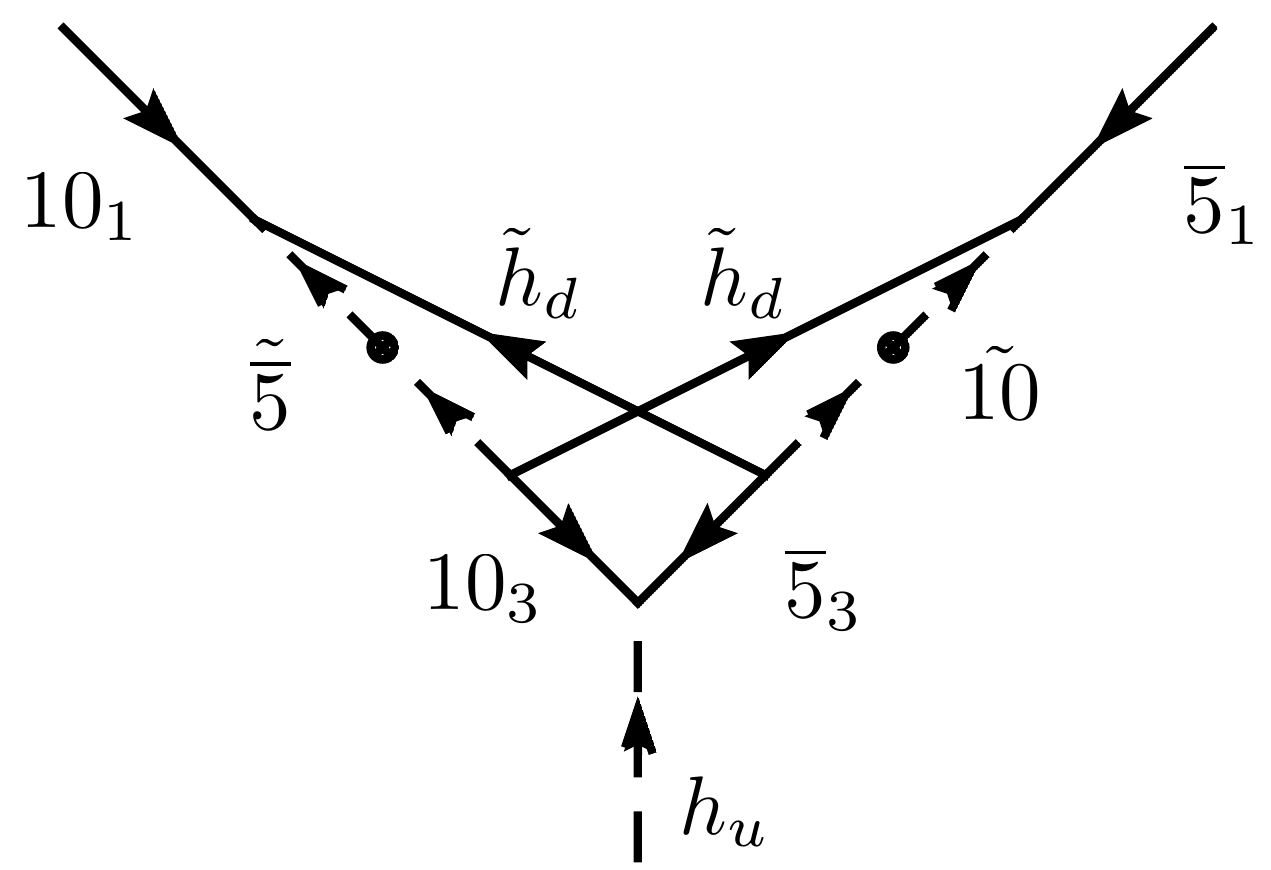}
\caption{\label{fig-scalarfirstgen} Squark and slepton mixing leads to first generation masses at two loops below the top quark mass.  }
\end{center}
\end{figure}

SUSY breaking  in the form of mass splittings between the fermion and scalar components of  $H_u, \, H_d$ and a $\tripletbmu \hutriplet \hdtriplet$ term for the triplet higgs scalars is sufficient to generate all the quark and lepton masses (see sections \ref{UpMasses} and \ref{DownMasses}). We will first assume that these are the only tree-level SUSY breaking operators and analyze their effects on the mass generation in the up sector ({\it i.e.} the $10_i \otimes 10_j$ sector). The up quarks $U_1$ and $U_2$ couple to the flavorful part of the superpotential (\ref{UglyPotential}) through the higgs triplets. The loops responsible for their masses will involve the higgs triplets and these loops will be dominated by momenta of order the triplet masses $\sim \Mgut$. SUSY breaking $\sim \Mgut$ in this sector leads to scalar masses $m^2_{S}$ and mixings $\delta m^2_S$  one loop below the GUT scale $ \sim \loopfactor  \l \lambdac \r^2 \Mgut^2$. Since the generated masses are all of comparable order, the scalars are maximally mixed with $\OrderOne$ mixing angles. These mixing angles can induce masses for the first generation at two loops through the diagrams in \ref{fig-scalarfirstgen}. Due to GIM cancellations that occur at momenta much larger than the scalar masses, these two diagrams yield finite threshold corrections  $\propto  \l  \frac{\delta m^2_S}{m^2_{\hdtriplet}} \r^2$ (where the square arises from the need to insert two mixing angles in order to generate a mass for the first generation). As discussed above, the scalar mixings are a loop factor below the triplet masses and hence this contribution is parametrically of the same order as the four loop contributions computed earlier. Similarly, the contributions of these scalar masses to mixing between the first generation and the other generations is also suppressed. The hierarchies in the up quark sector are therefore parametrically unaffected by SUSY breaking in the higgs sector.

The generation of down and lepton masses involves the exchange of higgs doublets, in particular, the light higgsinos. In this case, the threshold corrections to the first generation masses discussed above are $\propto \l \frac{\delta m^2_S}{m^2_S} \r ^2$. This correction is parametrically of the same order as the previously computed five loop contribution only if $\delta m^2_S \sim \loopfactor m^2_S$. This can be achieved if the scalars receive flavor diagonal SUSY breaking masses $m^2_S \sim \Mgut^2$. Thus, a SUSY breaking mechanism that generates comparable flavor diagonal scalar and higgs sector masses will parametrically generate a mass hierarchy. 

Note that a parametric separation between the first and second generation masses is not necessary. The threshold corrections computed above are not enhanced by the ubiquitous $\log$ factor in the previously computed $\log$ divergent diagrams. Furthemore, these contributions are proportional to the fourth power of couplings and hence even a modest hierarchy $\sim \frac{1}{3}$ in the couplings can suppress these contributions without the need for flavor diagonal scalar masses.  In this case, the contributions from this finite threshold correction will be numerically comparable to the parametric five loop contributions.


\subsubsection{The Final Mass Matrix}
\label{FinalMass}
The yukawa couplings $y_{ij} 10_i 10_j H_u$ and $x_{ij} 10_i \FiveBar_j \HuDagger$ generated by the domino mechanism have the following parametric form
\begin{equation}
Y \sim \left(\begin{matrix} \epsilon^4 & \epsilon^4 & \epsilon^4 \\ \epsilon^4 & \epsilon^2 & \epsilon^2 \\ \epsilon^4 & \epsilon^2 & 1 \end{matrix}  \right), \qquad X \sim \left(\begin{matrix} \delta \epsilon^4 & \delta \epsilon^4 & \delta \epsilon^4 \\ \delta \epsilon^3 & \delta \epsilon^2 & \delta \epsilon^2 \\ \delta \epsilon^2 & \delta \epsilon & \delta \end{matrix}  \right)
\label{Eqn-yukawamatrices}
\end{equation}
where $\epsilon$ and $\delta$ parametrize the cost of each insertion of the spurions $\lambda \lambda^{\dagger}$ and $\lambda^{\dagger}$ respectively (see section \ref{Spurions}). In addition to these yukawa couplings, the diagrams discussed in this model also generate wave-function renormalizations $z^{ij}_{10}\TenDagger_i 10_j$ and $z^{ij}_{\FiveBar} \FiveBarDagger_i \FiveBar_j$ that are parametrically of the form 
\begin{equation}
Z_{\FiveBar} \sim Z_{10} \sim \left(\begin{matrix} 1 & \epsilon & \epsilon^2 \\ \epsilon & 1 & \epsilon \\ \epsilon^2 & \epsilon & 1 \end{matrix}  \right)
\label{Eqn-Zmatrices}
\end{equation}
Upon diagonalization of the kinetic terms, we get the mass matrices, 
\begin{equation}
Y_{u} \sim \left(\begin{matrix} \epsilon^4 & \epsilon^3 & \epsilon^2 \\ \epsilon^3 & \epsilon^2 & \epsilon \\ \epsilon^2 & \epsilon& 1 \end{matrix}  \right) \qquad \text{and} \qquad Y_d \sim \left(\begin{matrix} \delta \epsilon^4 & \delta \epsilon^3 & \delta \epsilon^2 \\ \delta \epsilon^3 & \delta \epsilon^2 & \delta \epsilon \\ \delta \epsilon^2 & \delta \epsilon & \delta \end{matrix}  \right)
\label{Eqn-yukawadownmatrices}
\end{equation}
for the up  $\l Y_u \, 10 \, 10\, H_u\r$ and down $\l Y_d\, 10\, \FiveBar\, \HuDagger\r$ sectors respectively. The parametric form of these mass matrices ({\it i.e.} the masses and mixing angles) are identical to the predictions of the spurion analysis of  section \ref{Spurions}. 

The domino mechanism has generated these mass matrices by repeated use of two spurions, $\lambda$ and $c$. After phase rotations, $\lambda$ can be made into a real matrix with 5 independent parameters, 3 of which generate the second and third generation masses. Phase rotations can also be used to make $c$ real, yielding one additional parameter.  The mechanism also requires a $\tripletbmu$ term to communicate chiral symmetry breaking to the down and lepton sector. Together, these 5 parameters ({\it i.e.} 3 from $\lambda$, 1 from $c$ and 1 from  $\tripletbmu$) generate the 5 parameters of the second and third generation masses (2 up quark masses, 2 down and lepton masses and one mixing angle).  The number of input parameters is equal to the number of measured Yukawa couplings, so we do not have exact predictions.  However, the input parameters have a very limited range of acceptable values, all must be $\OO(1)$, and so all the Yukawas are predicted up to $\OO(1)$ factors.  Hence, the domino mechanism can explain the hierarchies in the observed masses and mixing angles. 

In addition to these, the mass matrices also receive contributions from Planck suppressed operators (see section \ref{PlanckSlop}). The contributions from these operators $\sim \mathcal{O}\l\epsilon^3\r - \mathcal{O}\l\epsilon^4\r $ are significant only for the first generation mass and its mixing with the second generation. Furthermore, these operators can also introduce CP violating phases into the theory. The Planck suppressed operators that involve only the first and second generation will induce at least four new parameters including two phases in the up quark sector. Similarly, Planck suppressed operators can also contribute masses, mixings and phases to the first generation down sector masses. In addition to these operators, the down sector masses can also be affected by the vev $\hdvev$ of the doublet $h_d$. With all these parameters that contribute at levels comparable to the masses of the first generation, it is easy to accommodate the observed first generation masses in this framework. The numerical comparison between these matrices and the observed quark and lepton masses, including GUT breaking effects,  is performed in section \ref{PureSplitNumbers}. 

\subsubsection{Neutrino Masses}
\label{Sec: fives with neutrinos}

We have not attempted to explain the neutrino masses and mixings since they are already well described by either the standard seesaw mechanism or extra-dimensional mechanisms.  Following our general philosophy, the neutrino masses can be generated with appropriate values simply by allowing the terms $\mathcal{W} \supset H_u \, \FiveBar \, N + N \, N$ with arbitrary, flavor-blind coefficients (often called neutrino anarchy \cite{Hall:1999sn, Haba:2000be, deGouvea:2003xe}).  In our minimal model we may need to take the neutrino Yukawa couplings to be somewhat small and a correspondingly lighter $N$ mass (perhaps even at the TeV scale) in order to avoid feeding new flavor structure back into the quark and charged lepton sectors.  Such a structure can easily be generated in an extra-dimensional model.  For example, if all the SM fields are located on one brane (or localized in one place) in the extra-dimension, except the right handed neutrino field $N$, the Yukawa couplings are naturally small.  In fact, such a scenario could also easily give appropriate Dirac neutrino masses through the exponential suppression of the neutrino Yukawa couplings \cite{Agashe:2008fe}.  Since neutrino masses and mixings may not have a structure and may in fact be anarchic, we do not attempt here to explain them further with our domino mechanism.  This could be an interesting direction to explore further, especially since such a model might allow new possibilities for explaining the quark and charged lepton Yukawas as well.

It may also be possible to use the same neutrino mass model as proposed for our model with 45's (see Section \ref{Neutrino45}).  In this case though, the couplings $y_\nu$ in Eqn.~\eqref{eqn: neutrino 45 model} would need to be $\lesssim 10^{-2}$ instead of $\OO(1)$ to avoid feeding new flavor structure back into the quark and charged lepton Yukawas.

\subsubsection{Triplet Higgsinos}
\label{SplitTriplets}
In split SUSY, gaugino and higgsino masses are decoupled from scalar masses due to an unbroken $U(1)_R$ symmetry at low energies. This $R$ symmetry forbids $\mu \, H_u H_d$. Masses for the doublet higgsinos are generated when the $R$ symmetry is spontaneously broken at the $\sim$ TeV scale. However, the triplet higgsinos cannot be present at low energies since they ruin gauge coupling unification. Split SUSY requires a mechanism to eliminate the triplet higgsinos from the low energy spectrum without breaking the $R$ symmetry. In the model discussed in this section, the triplets play an essential role in generating fermion masses since they break the flavor symmetries of the $U$ quarks. However, note that all these symmetries are broken individually by both the scalar and fermion triplets. Consequently, quarks and lepton masses will be generated in this mechanism in the presence of the scalar triplets $\hutriplet$ and $\hdtriplet$. 

These two observations suggest a possible UV completion that can incorporate this domino mechanism in split SUSY. Consider a scenario with GUT scale extra dimensions in  which the triplet higgsino zero modes are projected out using orbifolds, while retaining the zero modes of the scalar triplets. This orbifold breaks SUSY at the compactification scale $\sim \Mgut$. The elimination of the triplet higgsino zero modes by orbifolding eliminates the triplet higgsinos from the low energy spectrum without breaking the $R$ symmetry necessary to have light gauginos. The retention of the zero modes of the scalar triplets allows the domino mechanism to generate masses for all the quarks and leptons. This strategy of eliminating the higgsino triplets while retaining the scalar triplets and an unbroken $R$ symmetry leads to a consistent effective field theory with broken supersymmetry. Consequently, in addition to the UV completion suggested above using orbifold compactifcations, there may be other UV completions that incorporate these ideas.

\subsection{Model with 45's} 
\label{Model45}
In this section, we discuss the case where $\phibar$ is a $\FortyFiveBar$ of $SU(5)$. The mass generation mechanism proceeds through the superpotential  
\begin{equation}
\mathcal{W} \supset y_3 \, 10_3 \, 10_3 \,  H_u  + \lambdac \,10_i \, \FiveBar_j \, \phibar
\label{SuperPotential45}
\end{equation}
that is similar in spirit to the minimal model described in section \ref{Model5}, where  $\phibar$ was a $\FiveBar$. The key features of the model, including the repeated use of the flavor breaking spurions $\lambda$ and $c$ in (\ref{SuperPotential45}), are similar to those of the minimal model. Since the $\FortyFiveBar$  breaks baryon and lepton number, it  mediates proton decay. Consequently, as in section  \ref{Model5}, this model also requires split SUSY with SUSY broken at the GUT scale in order to generate the flavor mass hierarchy. As we will see, the choice of $\phibar$ as a $\FortyFiveBar$ ameliorates some of the weaknesses of the minimal model, including the need to tune $\hdvev$. 

We first discuss the generation of the top yukawa coupling, after which we sketch the radiative mass generation mechanism for the remaining quarks and leptons. We then point out that unlike the minimal model,  this model does not require additional tunings in the $B_{\mu}$ sector. The effects of planck suppressed operators are then addressed. We then comment on the possible UV completions of this model into the framework of split SUSY. The generation of neutrino masses in this framework is then discussed before concluding with a comparison of this model and the minimal model discussed in section \ref{Model5}. 


\subsubsection{The Top Yukawa}

The top yukawa can be generated in a flavor democratic way by only allowing $H_u$  (through choice of  $U(1)_H$ symmetry) to couple to a sector of the theory that has linear couplings to the $10_i$ fields (see section \ref{TopYukawa}). To that end, we introduce one new vector-like $10$ field $10_{N_1}$ and a singlet $\sigma$ whose vev $\sigmavev$ breaks the $U(1)_H$ symmetry that forbids fermion masses. We add the terms 

\begin{eqnarray}
\mathcal{W}  \supset  a_{11} \, 10_{N_1} \, 10_{N_1} \, H_u \,  +  \, c_{i}\,  \TenBar_{N_1}\, 10_{i}\, \sigma \,  + \, M_{1} \, 10_{N_{1}} \, \TenBar_{N_{1}}
\label{Eqn:Potential45}
\end{eqnarray}
to the superpotential (\ref{SuperPotential45}) instead of the yukawa coupling  $y_3 \, 10_3 \, 10_3 \, H_u$.  Integrating out the $10_{N_1}$, we get the effective operator

\begin{equation} 
\frac{a_{11}^2 \sigma^2 \l c_i 10_i \r \l c_j 10_j \r H_u}{M_{1}^2}
\label{top45eq1}
\end{equation}
generated by the diagram in figure \ref{fig-minfrog}. Choosing a basis in which the $10_3$ direction points along $c_i$ (see section \ref{TopYukawa}), we generate a yukawa coupling $y_3 = \l \frac{a_{11} \sigmavev}{M_1}\r^2$ for the top quark in a flavor democratic way.

\subsubsection{The Mass Hierarchy}
\label{MassHierarchy}

With the generation of the top quark mass, upon SUSY breaking, the flavor breaking term $\lambdac \, 10_i \, \FiveBar_j \, \phibar$ will cause masses for the charm and up quark at two and four loops below the top quark mass respectively. The generation of these masses is similar to the mechanism discussed  in section \ref{UpMasses} through two loop diagrams similar to  figure \ref{fig-2loopup}, with $H_d$ replaced by $\phibar$. 

Down quark and lepton masses can be generated only after the breaking of the chiral symmetry in the $10_i \otimes 10_j$ space is communicated to the $10_i \otimes \FiveBar_j$ space. This can be achieved by introducing $\phi$, a $45$ of $SU(5)$ that will communicate between the top quark and the flavor breaking spurion $\lambda$. We will also add a new vector-like $10$ field $10_{N_{2}}$ to the theory. The need for this new field will become apparent shortly. To the superpotential (\ref{Eqn:Potential45}), we add the terms 

\begin{eqnarray}
\mathcal{W}  \supset   \, a_{12}\,  10_{N_1}\, 10_{N_2}\, H_u +  \, b_{12}\,  10_{N_1}\, 10_{N_2}\, \phi  \, + \, M_{2}\, 10_{N_{2}}\, \TenBar_{N_{2}}
\label{Eqn:Potential45num2}
\end{eqnarray}

\begin{figure}
\begin{center}
\subfigure[]{\label{fig-minfrog45}}
\includegraphics[width=2.5 in]{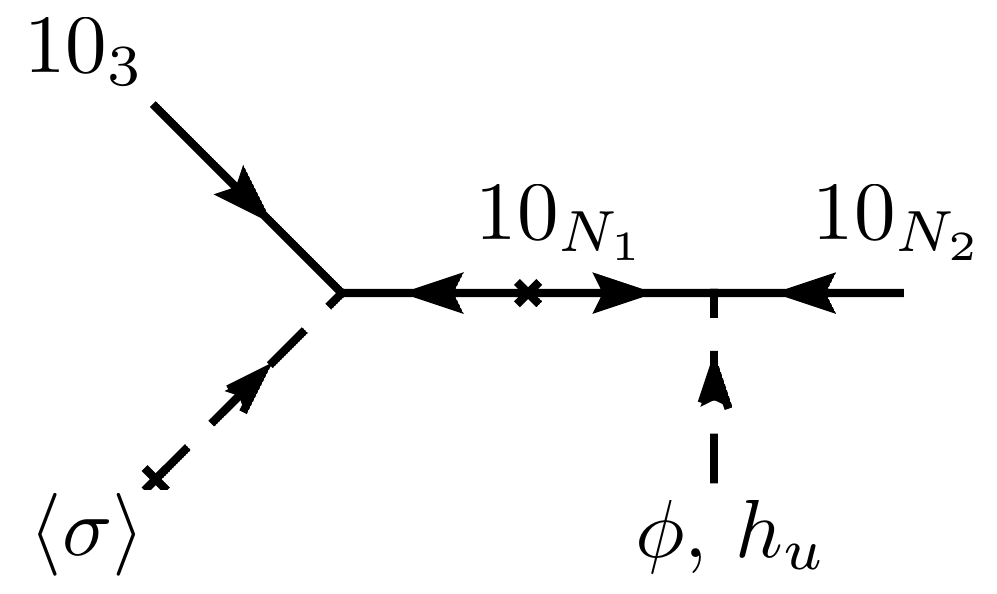} \qquad \qquad \qquad \qquad
\subfigure[]{\label{fig-1loopbtau45}}
\includegraphics[width=2.1 in]{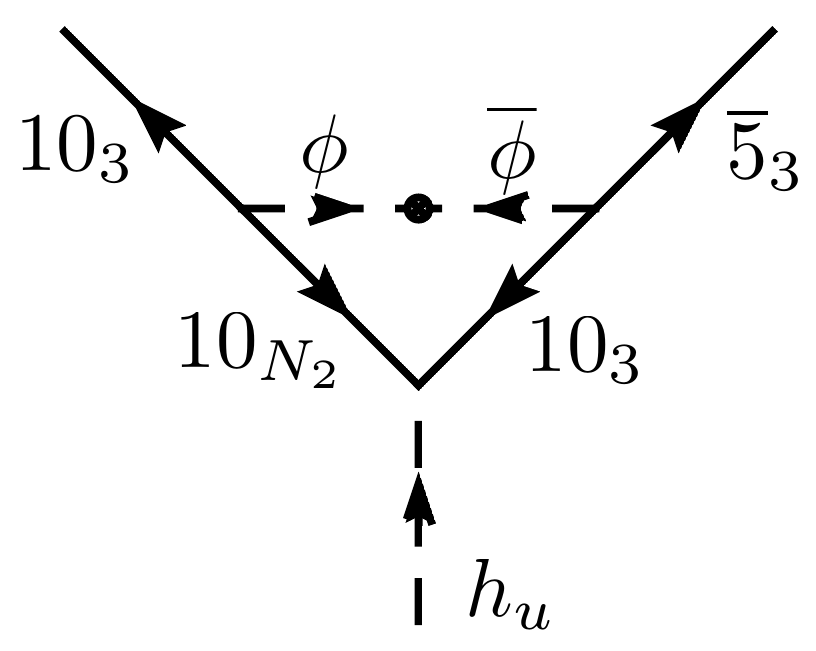}
\caption{ Figure \ref{fig-minfrog45} shows the generation of the operators in Eqn.~\eqref{45topcouplings} by integrating out the $10_{N_{i}}$ fields.
Figure \ref{fig-1loopbtau45} generates a $b-\tau$ mass $10_3 \, \FiveBar_3 \, \huDagger$ at one loop below the top mass, using the operators in Eqn.~\eqref{45topcouplings} and the SUSY breaking $B_{\mu} \, \tilde{\phi} \, \tilde{\phibar}$ term.}
\end{center}
\end{figure}

We introduced two $10_{N_i}$ fields since $\phi$, a $45$ of $SU(5)$, couples to the antisymmetric combination of two $10$s, which is zero unless the two $10$s are different fields.  Integrating out the $10_{N_i}$ fields, we get the effective operators 

\begin{equation} 
 \frac{a_{12} \sigma \l c_i 10_i \r 10_{N_{2}} H_u}{M_{1}} \, \text{ and } \, \frac{b_{12} \sigma \l c_i 10_i \r 10_{N_{2}} \phi}{M_{1}}
\end{equation}
generated by the diagrams in figure \ref{fig-minfrog45}. Choosing a basis in which the $3$ direction points along $c_i$, once $\sigma$ gets a vev $\sigmavev$, we generate the following interactions for the top quark:

\begin{equation} 
\l\frac{a_{12} \sigmavev}{M_{1}}\r \, 10_3 \, 10_{N_{2}} \, H_u \, \text{ and } \, \l\frac{b_{12} \sigmavev} {M_{1}}\r \, 10_3 \, 10_{N_{2}}\, \phi
\label{45topcouplings}
\end{equation}

In the presence of a SUSY breaking $B_{\mu} \, \tilde{\phi} \, \tilde{\phibar}$ term, these interactions will communicate the breaking of the chiral symmetry in the $10_i \otimes 10_j$ sector to the flavor-breaking $10_i \otimes \FiveBar_j$ sector. With these, the one loop diagram of figure \ref{fig-1loopbtau45} will generate a $b-\tau$ mass $x_3 \, 10_3 \,\FiveBar_3 \, \huDagger$. Once chiral symmetry is broken by this $b-\tau$ mass in the $10_i \otimes \FiveBar_j$ space,  $s-\mu$ and $d - e$ masses are generated at two and four loops below the $b-\tau$ mass. This mechanism is similar to the one discussed in section \ref{DownMasses} and operates through two loop diagrams similar to figure \ref{fig-2loopdown} with $H_d$ replaced by $\phibar$. 

$U(1)_H$ charge assignments that allow the operators in (\ref{Eqn:Potential45}) and (\ref{Eqn:Potential45num2}) but not the standard model yukawa couplings $10_i \, 10_j \, H_u$ and $10_i \, \FiveBar_j \, H_d$ are shown in Table \ref{Tab:charges45}.  The $\phi$, $\phibar$ fields cannot exist at low energies since they spoil gauge coupling unification. While the scalars  $\tilde{\phi}$ and $\tilde{\phibar}$ can receive SUSY breaking masses at the GUT scale, the elimination at low energies of the fermionic $\phi$ and $\phibar$ must be achieved without breaking the $R$ symmetry necessary to have light gauginos. We discuss these issues that arise in split SUSY in section \ref{Split45}. 

Note that the superpotential (\ref{Eqn:Potential45}) contains the term $c_i \, \TenBar_{N_{1}} \, 10_i \, \sigma$ but not  $d_i \, \TenBar_{N_{2}} \, 10_i \, \sigma$. The domino mechanism requires the absence of this term as it picks out another direction in flavor space, upsetting the hierarchy between the first and second generation masses (see section \ref{Spurions}). However, since we need the operators $a_{11} \, 10_{N_{1}} \, 10_{N_{1}} \, H_u$, $a_{12} \, 10_{N_{1}} \, 10_{N_{2}} \, H_u$ and $c_i \, \TenBar_{N_{1}} \, 10_i \, \sigma$ to generate (\ref{45topcouplings}), it is impossible to find $U(1)_H$ charge assignments that will allow $c_i \, \TenBar_{N_{1}} \, 10_i \, \sigma$ but not  $d_i \, \TenBar_{N_{2}} \, 10_i \, \sigma$. Since these are superpotential terms, as an effective field theory, it is consistent to have one without the other due to SUSY non-renormalization theorems. A simple UV completion that contains all the terms of (\ref{Eqn:Potential45}) but a highly suppressed $d_i \, \TenBar_{N_{2}} \, 10_i \, \sigma$ can be realized using locality in extra-dimensional theories. For example, in a brane-world scenario, the superpotential (\ref{Eqn:Potential45}) can be realized by localizing  $\TenBar_{N_1}$ and $10_i$ at one brane and $\TenBar_{N_2}$ at the other brane  with all the other fields having flat profiles in the bulk. We will not explore these UV completions further in this paper and will instead concentrate on the phenomenology of  (\ref{Eqn:Potential45}). 

The operators in (\ref{Eqn:Potential45}) also cause kinetic mixing between generations through wavefunction renormalizations. The dominant one loop diagrams that contribute to these mixings are similar to the one loop diagrams for the minimal model discussed in section \ref{Zs}, with the $H_d$ fields in those diagrams replaced by $\phibar$. The masses and kinetic mixings generated by this mechanism are parametrically similar to the mass matrices for the minimal model discussed in section \ref{FinalMass}. Upon diagonalization of the kinetic terms, these matrices yield hierarchial quark and lepton mass matrices. The generated hierarchies are parametrically identical to the predictions of the spurion analysis presented in section \ref{Spurions}.

\subsubsection{The Absence of Tuning}
\label{Tuning}

In addition to the radiative contributions to the fermion masses discussed above, a vev $\phibardoubletvev$ for the doublet component $\phibardoublet$ of $\phibar$ directly causes quark and lepton masses through the term $\lambdac \, 10_i \, \FiveBar_j \,  \phibar$. Similarly, a vev $\phidoubletvev$ for the doublet component of $\phidoublet$ of $\phi$ also causes fermion masses since $\phidoubletvev \sim \phibardoubletvev$ as a result of the large $B_{\mu} \tilde{\phi} \tilde{\phibar}$ term. In the absence of tuning, these contributions must not be larger than the first generation masses and hence we need $\phibardoubletvev, \, \phidoubletvev \lessapprox \mathcal{O}\l10^{-5}\r \huvev$. Since electroweak symmetry is broken by $\huvev$, $\phidoubletvev$ and $\phibardoubletvev$ are generated through operators of the form  $\tildephidoublet h_{u}^{*}$ and $\tildephibardoublet h_u$.

These operators can be generated only after $SU(5)$ is broken. It is conceivable that the physics responsible for breaking $SU(5)$ generates these operators without any suppression. The construction of a UV theory that incorporates the breaking of $SU(5)$ is beyond the scope of this paper. However, we will now present some examples of $SU(5)$ breaking where these terms are sufficiently suppressed.

$SU(5)$ could be broken through a vev  $\Sigmavev$ for $\Sigma$, a $24$ of $SU(5)$. In this case, the operator  $\tildephidoublet h_{u}^{*}$ is extracted from the $SU(5)$ invariant operator $\Sigma^{\dagger} \, \Sigma \, \phi^{\dagger} \, H_u$ after SUSY and $SU(5)$ are broken. This dimension six operator is the first operator that allows for the extraction of  $\tildephidoublet h_{u}^{*}$ since lower dimension operators that involve one $\Sigma$ field insertion can be easily forbidden by choice of $U(1)_H$ charge assignments for $\Sigma$. The generation of this diagram requires gauge interactions since in the absence of gauge interactions, $\Sigma$ does not know about the existence of $\phi$ or $H_u$. The $10_{N_{i}}$ sector must also be involved in generating this operator since in the absence of these fields, there exist $U(1)_H$ charge assignments that forbid $\Sigma^{\dagger} \, \Sigma \, \phi^{\dagger} \, H_u$. The diagram that generates this operator must therefore involve at least two loops since it must know about both  the $10_{N_{i}}$ sector and gauge interactions. The contribution from this diagram is  suppressed by at least $\sim \l \loopfactor \r^2 g^4 $. Furthermore, since this is a dimension 6 operator, a mild hierarchy between $\Sigmavev$ and the cut-off will also rapidly suppress this operator. These considerations lead to  $\phidoubletvev \lessapprox  \mathcal{O}\l10^{-5}\r \huvev$. Similarly, we also have $\phibardoubletvev \lessapprox  \mathcal{O}\l10^{-5}\r \huvev$. Consequently, in this model, SUSY breaking effects lead to fermion mass contributions comparable to that of the first generation, without the need for additional tuning. 

Another way to break $SU(5)$ while eliminating these operators is to project out the doublet components $\phidoublet$ and $\phibardoublet$ using an orbifold. The remaining multiplets in the $45$ also break all the flavor symmetries and the domino mechanism will generate hierarchial quark and lepton masses through them.

\begin{table}
\begin{center}
\begin{math}
\begin{array}{|c|c|c|}
\hline
\text{Field} & U(1)_H \text{ Charge}  & \text{Example} \\
\hline
\sigma & 1 & +1 \\
\hline
10_i &  Q_{10} & +1 \\
\hline
H_u, \phi & -2 \l 1 + Q_{10}\r & -4 \\
\hline
\phibar & 2 \l 1 + Q_{10} \r & +4 \\
\hline
\FiveBar_i & -\l 2 + 3\, Q_{10} \r  & -5 \\
\hline
10_{N_{1}}, 10_{N_{2}} & \l 1 + Q_{10}\r&-2\\
\hline
H_d &  Q_d & +2 \\
\hline
\end{array}
\end{math}
\caption[Example Charge Assignments]{\label{Tab:charges45}  $U(1)_H$ charge assignments that allow the terms in (\ref{Eqn:Potential45}), (\ref{Eqn:Potential45num2}) and $B_{\mu} \,  \tilde{\phi} \, \tilde{\phibar}$ but not the yukawa couplings $10_i \, 10_j \, H_u$, $10_i \, \FiveBar_j \, H_d$.}
\end{center}
\end{table}

\subsubsection{Planck Slop}
\label{Planck45}
The masses of the first generation quarks and leptons can  also receive significant contributions from Planck suppressed operators (see section  \ref{PlanckSlop}). Any $U(1)_H$ charge assignment that allows the operators in Eqn.~\eqref{45topcouplings} will also allow the higher dimension operators 

\begin{equation}
\int d^2 \theta \frac{\sigma^2 H_u 10_i 10_j }{M^2_{pl}}
\end{equation}
These operators will contribute to the masses, mixings  and phases of the up quark after $\sigma$ acquires a vev $\sigmavev$. $U(1)_H$ charges $Q_d$ and $Q_{\phibar}$ of $H_d$ and $\phibar$ can be chosen to allow the  operators

\begin{equation}
\int d^2 \theta \frac{\sigma^2 H_d 10_i \FiveBar_j }{M^2_{pl}}, \qquad \int d^4 \theta \frac{\sigma^{\dagger} \HuDagger 10_i \FiveBar_j }{M^2_{pl}}
\label{PlanckSlop45}
\end{equation}
An example charge assignment that allows the first operator is shown in  Table \ref{Tab:charges45}. As we will see in section \ref{Split45}, this model can allow $\hdvev \sim \huvev$. In this case, once $\sigma$ acquires a vev $\sigmavev$, the first operator in  (\ref{PlanckSlop45}) can make significant contributions to the masses, phases and mixings of the $d$ quark and electron. The second operator in (\ref{PlanckSlop45}) was discussed earlier in section \ref{PlanckSlop} and gives rise to first generation masses and phases after $\sigma$ develops a SUSY breaking $F$ term vev $\langle F_{\sigma} \rangle$. This operator contributes even when $\hdvev \ll \huvev$.

\subsubsection{$\phi$, $\phibar$ Fermions}
\label{Split45}
The fermionic components of $\phi$, $\phibar$ cannot be present at low energies since they ruin gauge coupling unification. In this model, they can be removed in two ways while retaining unification. One way to achieve this goal is to follow the strategy outlined in section \ref{SplitTriplets} whereby a SUSY breaking orbifold is used to project out the zero modes of the fermionic  $\phi$ and $\phibar$ fields while retaining the scalar components $\tilde{\phi}$ and $\tilde{\phibar}$. In this scheme, the elimination of the fermionic $\phi$ and $\phibar$ zero modes is done without breaking the $R$ symmetry that protects gaugino and doublet higgsino masses, leading to the familiar split SUSY scenario with gauginos and higgsinos of mass $\sim$ TeV.

This model also allows for another possibility. Split SUSY required the existence of gauginos and higgsinos at low energies in order to allow unification. However, as pointed out in \cite{MinimalSplit}, the gauge couplings unify even if the low energy spectrum contained only light doublet higgsinos. A spectrum with $\sim$ TeV scale doublet higgsinos and gaugino masses at the SUSY breaking scale is radiatively stable in the limit $\hdvev \ll \huvev$ ({\it i.e.} the $\tan \beta \rightarrow \infty$ limit) \cite{Giudice:2004tc}. In this limit, the higgsinos can be light since their masses are protected by a $U(1)_{PQ}$ symmetry. The model discussed in this section is compatible with  this symmetry  since the model does not require large $PQ$ breaking operators (unlike the $\tripletbmu \hutriplet \hdtriplet$ term in the model discussed in \ref{Model5}). Furthermore, since the mass generation only involves $\huvev$, this model also allows $\tan \beta \rightarrow \infty$. Consequently, in this model, the fermionic $\phi$ and $\phibar$ can receive $\mu_{\phi} \sim $ GUT scale masses through $R$ symmetry breaking operators of the form $\mu_{\phi} \, \phi \, \phibar$. With a broken $R$ symmetry, the gauginos become heavy and  the low energy spectrum of the model only contains the higgsino doublets, thus retaining unification.

\subsubsection{Neutrino Masses}
\label{Neutrino45}

Just as for the minimal model in Section \ref{Sec: fives with neutrinos}, the neutrino masses can be simply generated by either a standard seesaw or extra-dimensional mechanism.  In the model discussed in this section there is an additional possibility.  We suggest a way to add a radiative mechanism to the standard seesaw which makes new predictions for neutrino masses and mixings.  Instead of simply allowing the standard heavy right-handed neutrino mass term, we will forbid it and then generate it with a mechanism similar to that used for the quarks and charged leptons.  In addition to the usual three right handed neutrinos, $N_i$, we add one more which we call $\overline{N}$.  We charge the neutrino fields under our U(1) symmetry as in Table \ref{Tab:neutrinocharges45}.  So only $\overline{N}$ has an allowed mass term.  The other three $N_i$ will get mass when $\sigma$ gets a vev breaking the U(1) symmetry.

\begin{figure}
\begin{center}
\includegraphics[width=3 in]{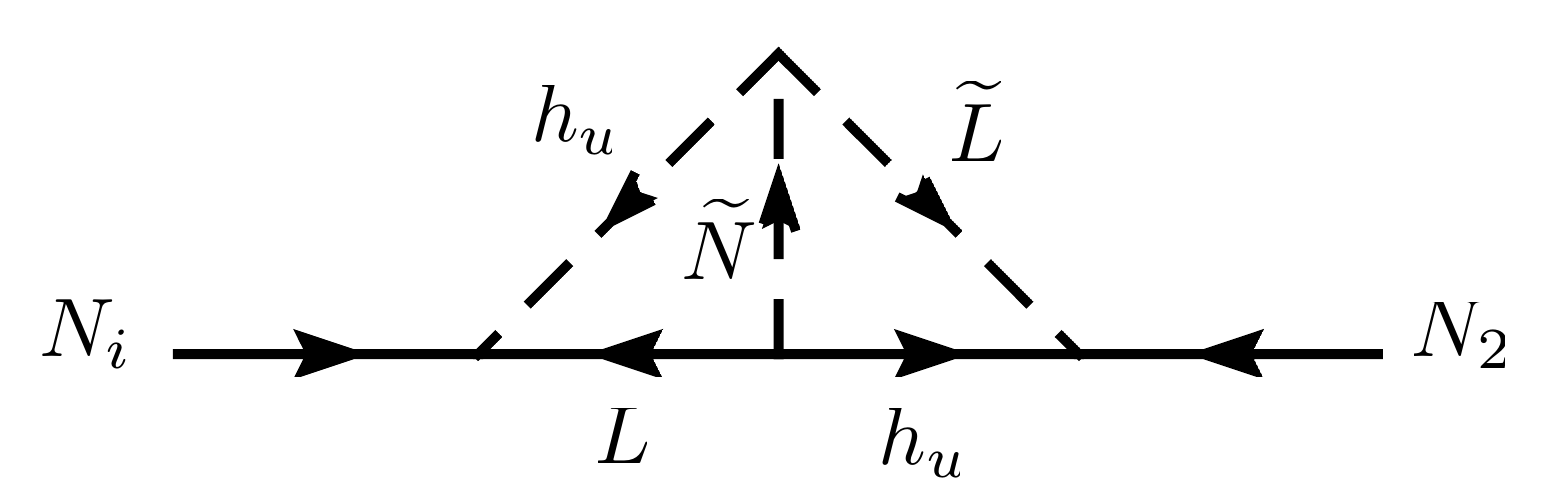}
\caption{ \label{fig: neutrinoloop} The diagram which generates the right handed neutrino masses for our neutrino model of Section \ref{Neutrino45}.  Note that we have labelled the neutrino line on the right as $N_2$ to illustrate that it is only a single direction in $N_i$ space and it is in general different from $N_1$.  By contrast the other neutrino is labelled $N_i$ to illustrate that it can be any of the 3 right handed neutrinos.}
\end{center}
\end{figure}

If we then write down all allowed terms in the superpotential
\begin{equation}
\label{eqn: neutrino 45 model}
\mathcal{W} \supset {y_\nu} \, H_u \, \overline{5}_i \, N _j+ c_i \, \sigma \, N_i \, \overline{N} + M_{\overline{N}} \, \overline{N} \, \overline{N}
\end{equation}
where all couplings are $\OO(1)$ and all scales ($\langle \sigma \rangle$ and $M_{\overline{N}}$) are near $M_\gut$.
There is a basis in which the coupling $c_i$ points only in the $N_1$ direction.  Thus when we integrate out the $\overline{N}$ it will generate a mass at tree level for only one of the right handed neutrinos, which we will choose to call $N_1$.  If we also have an arbitrary SUSY-breaking scalar mass matrix for the $\widetilde{N}_i$, then masses are generated for both the remaining $N_2$ and $N_3$ through the diagram in Figure \ref{fig: neutrinoloop}.  This generates a mass matrix for the $N_i$ of the form:
\begin{equation}
\label{Eqn right neutrino mass}
M_N \propto M_\gut \left(\begin{matrix} 1 & \epsilon^2 & 0 \\ \epsilon^2 & \epsilon^2 & \epsilon^2 \\ 0 & \epsilon^2 & 0 \end{matrix}  \right)
\end{equation}
where $\epsilon^2$ stands for the two-loop diagram in Figure \ref{fig: neutrinoloop}.  This texture arises from the choice of direction $N_2$ as explained in that figure.  Notice that we have generated one right handed neutrino mass at the GUT scale and two that are a couple orders of magnitude below the GUT scale.  This generates a Majorana neutrino mass $\left( H_u L \right)^\intercal \, y_\nu \, M^{-1}_N \, y^\intercal_\nu \, \left( H_u L \right)$.  Because of the spectrum of right handed neutrino masses, two of the Majorana neutrino masses are heavier while one is lighter by a couple orders of magnitude, the `inverted hierarchy.'  This predicts the correct order of magnitude for the neutrino masses and the arbitrary matrix $y_\nu$ will in general lead to large mixing angles (though of course $\theta_{13}$ has to be a bit small $\lesssim 12^\circ$).  Thus we have naturally explained the size of the neutrino masses and mixings, including why (two of) the right handed neutrino masses are not at the GUT scale but are a couple orders of magnitude below.

\begin{table}
\begin{center}
\begin{math}
\begin{array}{|c|c|}
\hline
\text{Field} & U(1)_H \text{ Charge} \\
\hline
\sigma & +1 \\
\hline
10_i & -1 \\
\hline
H_u, \phi & 0 \\
\hline
\phibar & 0 \\
\hline
\FiveBar_i & +1 \\
\hline
10_{N_{1}}, 10_{N_{2}} & 0 \\
\hline
H_d & Q_d \\
\hline
N & -1 \\
\hline
\overline{N} & 0 \\
\hline
\end{array}
\end{math}
\caption[Example Charge Assignments]{\label{Tab:neutrinocharges45}  $U(1)_H$ charge assignments for our neutrino model of Section \ref{Neutrino45}.  This is different from the example shown in Table \ref{Tab:charges45} but still allows our model for the quark and charged lepton Yukawa couplings.  Note that for this model the charge of $10_i$ is fixed.  These charges allow only the terms in \eqref{eqn: neutrino 45 model}, though this also requires choosing the R-parity of $\overline{N}$ to be +. }
\end{center}
\end{table}

\subsubsection{Comparison}
The minimal model discussed in section \ref{Model5} generated the fermion mass hierarchy with the addition of just one new, GUT scale,  vector like $10$ to split SUSY. In this model, the entire flavor hierarchy was generated using the $H_d$ field. However, the model requires a tuning of the  $B_{\mu}$ terms so that $\frac{\doubletbmu}{\tripletbmu} \sim 10^{-5}$ (see section \ref{DownMasses}). This tuning is not required when $\phibar$ is a $\FortyFiveBar$ (see section \ref{Tuning}). 

Additionally, in the minimal model, since the doublet higgsinos are light, parametric separation between the first and second generation masses required flavor diagonal scalar masses (see section \ref{superpartner}). When $\phibar$ is a $\FortyFiveBar$, the fermionic components of $\phi$, $\phibar$ are either projected out or have GUT scale masses (see section \ref{Split45}). When the $\phi$, $\phibar$ have GUT scale masses,  a GIM-like cancellation exists in superpartner diagrams like figure \ref{fig-scalarfirstgen} as long as the squark and slepton masses are smaller than the fermionic $\phi$, $\phibar$. If the primary SUSY breaking source in the theory occurs in the $\phi$, $\phibar$ sector, the contributions from the squark and slepton masses generated at one loop from this SUSY breaking contribute to the first generation masses at parametrically the same order as the radiative mass generation mechanism.  This is analogous to the case of the higgs triplet discussed in section \ref{superpartner}. These superpartner diagrams do not exist if the fermionic $\phi$ and $\phibar$ are projected out. The only SUSY breaking necessary in this model to produce the fermion mass hierarchy are the SUSY breaking interactions in the $\phi$, $\phibar$ sector. Flavor diagonal scalar masses are not required in this model. However, direct SUSY breaking contributions to the squark and slepton masses should still be  smaller  ($\sim \loopfactor$) than the masses in the $\phi$, $\phibar$ sector so that $\lambdac$ continues to be the dominant source of flavor breaking in the theory.

Numerically, as we will see in sections \ref{PureSplitNumbers} and \ref{Model45Numbers}, the model with $\FortyFiveBar$ has larger color factors and can therefore more easily accommodate some of the observed mass hierarchies, in particular, the $\mu-s$ mass. The case with $\FortyFiveBar$ is however less minimal. It requires the addition of more particle multiplets (two vector like $10_{N_{i}}$ fields as opposed to one vector like $10_N$ field in the minimal model) as well as the absence of certain superpotential terms (for example, $d_i \, \TenBar_{N_{2}} \, 10_i \, \sigma$ in equation  (\ref{Eqn:Potential45num2})  that are allowed by symmetries).

\section{Numerical Results}
\label{Sec: numerical results}
In this section we discuss numerical fits of the models discussed in this paper to the observed quark and lepton masses. We evaluate the color factors and estimate the size of the loops that enter into the domino mass generation mechanism. After estimating the sizes of these contributions, we discuss the emergence of GUT breaking effects in the generated masses. Contributions (including phases) from Planck-suppressed operators are then addressed. We first discuss the numerics of the minimal split model (see section  \ref{Model5}) followed by the model with $\FortyFiveBar$.

\begin{table}
\begin{center}
\begin{tabular}{|c|c|c|}
\hline
Description &  Figure   &   Color Factor  \\
\hline
Up sector, two loop& \ref{fig-2loopup} & -4 \\
\hline
Down sector, two loop& \ref{fig-2loopdown}& -2 \\
\hline
$10$ wavefunction renormalization& \ref{fig-wavefunction10} & +2\\
\hline
$\FiveBar$ wavefunction renormalization& \ref{fig-wavefunction5} & +4\\
\hline
 One loop $b-\tau$ mass & \ref{fig-1loopbtau} & -3\\
\hline
 $b-\tau$ mass only with triplets & \ref{fig-tripletbtau} & Leptons -3, Quarks -2\\
\hline
3 loop contribution to $s-\mu$ & \ref{fig-3loop} & Leptons 6, Quarks 4\\
\hline
2 loop mixing & \ref{fig-2loopnew} & Leptons -6, Quarks -4\\
\hline
\end{tabular}
\end{center}
\caption[Color Factors]{\label{Tab:color5}  Color factors for the diagrams of the minimal model discussed in section \ref{Model5}. }
\end{table}

\subsection{Minimal Model}
\label{PureSplitNumbers}
The minimal split model accomplishes the generation of the quark and lepton masses through loops involving the higgs multiplets $H_u$ and $H_d$. The color factors for the diagrams responsible for this mass generation are listed in Table \ref{Tab:color5}. These color factors were evaluated for the superpotential 
\begin{equation}
\mathcal{W} \supset \frac{y_3}{8} \, 10_3 \, 10_3 \, H_u \, + \, \lambdac \,10_i \, \FiveBar_j \, H_d 
\label{dummy5}
\end{equation}
with canonical normalization of the kinetic terms. This superpotential is derived from
\begin{equation}
\mathcal{W} \supset \lambdac \, 10_i  \, \FiveBar_j \,  H_d \,+\, c_i\, \sigma \, 10_i \, \TenBar_N  \, + \,  M \, 10_N \, \TenBar_N \,+\, \frac{y_N}{8} \,10_N\, 10_N \, H_u
\end{equation}
which yields eqn.~\eqref{dummy5} with $y_3 = \l \frac{y_N \sigmavev}{M} \r^2$ after $10_N$ is integrated out and $\sigma$ acquires a vev $\sigmavev$ (see section \ref{TopYukawa}).  The SUSY breaking $B_{\mu}$ term responsible for transmitting the breaking of the chiral symmetry to the $10_i \otimes \FiveBar_j$ space was taken to be 
\begin{equation}
\mathcal{L}_{\text{\sout{SUSY}}} \supset  B_{\mu} \, h_u \,  h_d
\end{equation}

We now estimate the sizes of the loop diagrams described in Table \ref{Tab:color5}. The  diagrams in figures \ref{fig-2loopup} and \ref{fig-2loopdown} are $\log$ divergent.  The divergent piece of this diagram was computed using dimensional regularization. The value of the diagram depends upon color factors and couplings and evaluates to 
\begin{eqnarray}
\epsilon_2 \times \l \text{color factor} \r \times \l \text{couplings} \r
\end{eqnarray}
where the loop factor $\epsilon_2$ is 
\begin{eqnarray}
\epsilon_{2} & = & \l \loopfactor \r^{2} \log \l \frac{\Lambda^{2}}{m^{2}_{h_{d}}}\r,
\label{looplogfactor}
\end{eqnarray}
$\Lambda$ is the UV cut-off of the loops and $m_{h_d}$ is the mass of the triplet $h_d$ (see section \ref{UpMasses}) . 

Similarly, the one loop wave-function renormalization diagrams \ref{fig-wavefunction10} and \ref{fig-wavefunction5} are also $\log$ divergent. The size of the divergent part of the diagram is 
\begin{eqnarray}
\epsilon_{Z} & = & - \l \loopfactor \r \log \l \frac{\Lambda^2}{m^{2}_{h_{d}}} \r
\label{SizeZ}
\end{eqnarray}

The one loop diagram \ref{fig-1loopbtau} is convergent and its magnitude is 
\begin{eqnarray}
\delta & = & \frac{1}{2} \l \loopfactor \r \log \l \frac{m^{2}_{h_{d}} + B_{\mu}}{m^{2}_{h_{d}} - B_{\mu}}\r 
\end{eqnarray}

The topology of the three loop diagram in figure \ref{fig-3loop} is similar to that of the effective three loop diagram in figure \ref{fig-2loopdown} that yields the $s-\mu$ mass. Both diagrams have one non-planar loop, in the absence of which they yield planar, nested loops. The divergence structure of these diagrams should be similar and we estimate it to be 
\begin{eqnarray}
\sim \epsilon_2 \times \delta
\end{eqnarray}

The two loop diagram in figure \ref{fig-2loopnew} is a nested, planar diagram. The inner loop is very similar to the one loop wave-function renormalization described in figure \ref{fig-wavefunction10} while the outer loop is very similar to the one loop diagram responsible for generating the $b - \tau$ mass in figure \ref{fig-1loopbtau}. The divergences of this diagram factorize and their overall size is 
\begin{eqnarray}
\sim \epsilon_Z \times \delta
\end{eqnarray}

Using the color factors in Table \ref{Tab:color5}  and the values of the divergent contributions computed above, the leading parametric piece of the generated yukawa matrices $y_{ij} \, 10_i \, 10_j \, H_u$ and $x_{ij} \, 10_i \, \FiveBar_j \, \HuDagger$ are 
\begin{equation}
\begin{array}{rcl}
Y & = & y_3 \left(\begin{matrix} \minfourloopten \l \l -4 \r^2 \lambda_{12}^{2} \lambda_{22}^2 \lambda_{23}^2 \lambda_{33}^2 + \dots  \r & \minfourloopten \l \l-4\r^2 \lambda_{12} \lambda_{22}^3 \lambda_{23}^2 \lambda_{33}^2  + \dots \r  & \minfourloopten \l \l-4\r^2 \lambda_{12} \lambda_{22} \lambda_{23}^3 \lambda_{33}^3  + \dots \r \\  \\ \minfourloopten \l \l-4\r^2 \lambda_{12} \lambda_{22}^3 \lambda_{23}^2 \lambda_{33}^2  + \dots \r& \mintwoloopten \lambda_{23}^{2} \lambda_{33}^{2} & \mintwoloopten \lambda_{23} \lambda_{33}^{3} \\ \\ \minfourloopten \l \l-4\r^2 \lambda_{12} \lambda_{22} \lambda_{23}^3 \lambda_{33}^3  + \dots \r  & \mintwoloopten  \lambda_{23} \lambda_{33}^{3} & 1 - 4 \, \epsilon_2 \lambda_{33}^4 \end{matrix}  \right) \\ 
\end{array}
\end{equation}

\begin{eqnarray}
& &X  = -3 \, \lambda_{33} \, \delta  \, y_3 \times \nonumber \\ \nonumber \\
& &  \left(\begin{matrix}  \epsilon_2^2 \l 8 \, \lambda_{11}^2 \lambda_{12} \lambda_{22}^2 \lambda_{23}^2 \lambda_{33} + \dots \r  & \epsilon_2^2 \l 8 \, \lambda_{11}  \lambda_{22}^2 \lambda_{23}^3 \lambda_{33}^2 + \dots \r & \epsilon_2^2 \l 8 \, \lambda_{11}^2 \lambda_{12} \lambda_{22}^2 \lambda_{23}^2 \lambda_{33} + \dots \r  \\ \\ \epsilon_2 \epsilon_Z \l 4 \, \lambda_{22}^{2} \lambda_{33}^{2} \lambda_{12} \lambda_{11} + \dots  \r  & -4 \, \epsilon_{2} \lambda_{22} \lambda_{23}^2 \lambda_{33} & -4 \, \epsilon_{2} \l \lambda_{23}^3 \lambda_{33} + \lambda_{23} \lambda_{33}^3 \r  \\ \\ \epsilon_{Z}^2 \l 8 \, \lambda_{23} \lambda_{22} \lambda_{12} \lambda_{11} + \dots \r & -2 \, \epsilon_Z \lambda_{22} \lambda_{23} - 4 \, \epsilon_2 \lambda_{22}\lambda_{23} \lambda_{33}^2& 1 - 2 \, \epsilon_Z \l \lambda_{23}^2 + \lambda_{33}^2 \r - 4 \, \epsilon_{2} \lambda_{33}^2 \l \lambda_{23}^2 + \lambda_{33}^2 \r  \end{matrix}  \right) 
\label{Eqn-yukawamatrices5}
\end{eqnarray}

The first generation masses receive contributions from diagrams like figures \ref{fig-4loopup} and \ref{fig-5loopdown} in addition to the diagrams discussed in this section. These masses are also affected by Planck suppressed operators and the effects of SUSY breaking (see sections \ref{PlanckSlop} and \ref{superpartner}). 

The domino mechanism also generates wavefunction renormalizations. While these do not change the rank of the mass matrix, they can alter mixing angles and the magnitude of the generated masses. The leading parametric contributions to these renormalizations are estimated using equation \eqref{SizeZ} and the color factors from Table \ref{Tab:color5}. These yield 
\begin{eqnarray}
Z_{\FiveBar}  & = & \left(\begin{matrix} 1 - 4 \, \epsilon_Z \lambda_{11}^2 &  4 \, \epsilon_Z \lambda_{12} \lambda_{11} & 8 \, \epsilon_Z^2 \lambda_{11} \lambda_{12} \lambda_{22} \lambda_{23} \\ \\ 4 \, \epsilon_Z \lambda_{12} \lambda_{11} & 1 - 4 \,\epsilon_Z \l \lambda_{22}^2 + \lambda_{12}^2 \r& 4 \, \epsilon_Z \lambda_{23} \lambda_{22}  \\ \\ 8 \, \epsilon_Z^2 \lambda_{11} \lambda_{12} \lambda_{22} \lambda_{23} & 4 \, \epsilon_Z   \lambda_{23} \lambda_{22}  & 1 - 4 \epsilon_Z \l \lambda_{23}^2 + \lambda_{33}^2 \r \end{matrix}  \right), \\ \nonumber \\ \nonumber \\  Z_{10}  & = & \left(\begin{matrix} 1 - 2 \, \epsilon_Z \l \lambda_{11}^2 + \lambda_{12}^2\r &  2 \, \epsilon_Z \lambda_{12} \lambda_{22} & 8 \, \epsilon_Z^2 \lambda_{12} \lambda_{22} \lambda_{23} \lambda_{33} \\ \\ 2 \, \epsilon_Z \lambda_{12} \lambda_{22} & 1 - 2 \, \epsilon_Z \l \lambda_{22}^2 + \lambda_{23}^2 \r& 2 \, \epsilon_Z \lambda_{23} \lambda_{33}  \\ \\ 8 \, \epsilon_Z^2 \lambda_{12} \lambda_{22} \lambda_{23} \lambda_{33} & 2 \, \epsilon_Z   \lambda_{23} \lambda_{33}  & 1 - 2\, \epsilon_Z \lambda_{33}^2 \end{matrix}  \right)
\label{Eqn-ZmatricesNums5}
\end{eqnarray}
These wavefunction renormalizations are incorporated into the theory through the beta functions responsible for renormalizing the yukawa couplings $X$ and $Y$. The yukawa matrices $Y$ and $X$ get renormalized as 
\begin{eqnarray}
Y &  \rightarrow & \l \mathds{1} - \frac{1}{2} \l Z_{10} - \mathds{1} \r \r^{\intercal} Y  \l \mathds{1} - \frac{1}{2} \l Z_{10} - \mathds{1} \r \r \\
X &  \rightarrow & \l \mathds{1} - \frac{1}{2} \l Z_{10} - \mathds{1} \r \r^{\intercal} X  \l \mathds{1} - \frac{1}{2} \l Z_{\FiveBar} - \mathds{1} \r \r 
\end{eqnarray}
Upon diagonalization, the final mass matrix has the parametric form \eqref{Eqn-yukawadownmatrices}. 

We  numerically fit these diagonalized matrices to data. We take the Higgs vev to be $v = 174 \, \GeV$.  In the absence of a full calculation in split SUSY, we take the values of the Yukawa couplings at the GUT scale from those calculated in the SM \cite{Xing:2007fb, Fusaoka:1998vc, PDG} since the runnings are similar \cite{Giudice:2004tc, Arvanitaki:2004eu}.  Further, the Yukawa couplings at the GUT scale are similar in the SM and in the MSSM, so it seems likely that they will be similar in split SUSY.  Further, the values in split SUSY are likely to be closer to those in the SM than in the MSSM because of the lack of squarks and sleptons and the corresponding flavor directions they introduce.  So we take the values of the Yukawa couplings (after diagonalization) at the GUT scale to be
\begin{eqnarray}
Y^D_u & = &  \left( \begin{matrix} 3 \times 10^{-6} & 0 & 0 \\ 0 & 1.3 \times 10^{-3} & 0 \\ 0 & 0 & 0.4 \end{matrix} \right)
\nonumber \\
\label{actualdata}
Y^D_d & = &  \left( \begin{matrix} 7 \times 10^{-6} & 0 & 0 \\ 0 & 1.3 \times 10^{-4} & 0 \\ 0 & 0 & 6 \times 10^{-3} \end{matrix} \right)
\\
Y^D_l & = & \left( \begin{matrix} 3 \times 10^{-6} & 0 & 0 \\ 0 & 6 \times 10^{-4} & 0 \\ 0 & 0 & 10^{-2} \end{matrix} \right)
\nonumber
\end{eqnarray}
%
%
where the Yukawa couplings are $h_u \, Q \, Y^D_u \, U$, $h^\dagger_u \, Q \, V_\text{CKM} \, Y^D_d \, D$, and $h^\dagger_u \, L \, Y^D_l \, E$.  The CKM matrix is affected by running very little and we take the magnitudes to be
\begin{equation}
V_\text{CKM} \approx \left(
\begin{matrix}
1 & 0.2 & 0.004 \\
0.2 & 1 & 0.04 \\
0.009 & 0.04 & 1
\end{matrix}
\right)
\end{equation}
These are the values we will attempt to fit with our model.

The second and third generation quark masses and the mixing angle $V_{cb} = V_{ts}$ are reproduced with 
\begin{equation}
\begin{array}{lll} 
y_3 = \l \frac{y_N \sigmavev}{M}\r^2 \approx 1 \qquad & \frac{\tripletbmu}{m^{2}_{h_d}} \approx 1 \qquad & \log \l \frac{\Lambda^2}{m^{2}_{h_d}}\r \approx 4 \\ \\
\lambda_{22} \approx 1.6 \qquad & \lambda_{23} \approx 3 \qquad & \lambda_{33} \approx 0.6 \\ \\
\end{array}
\label{lambdaestimates}
\end{equation}
We note that these estimates were made by only including the divergent pieces of the diagrams discussed in Table \ref{Tab:color5}. Since the $\log$ factors used in these estimates are $\OrderOne$,  finite contributions to these diagrams are also signficant. Furthermore, these evaluations were made to two loop orders below the top mass for the up sector yukawa matrices and two loop orders below the $b-\tau$ mass for the down and lepton sector yukawa matrices. Higher order loops will correct these estimates and these corrections could also be significant since the estimated values of $\lambda$ in \eqref{lambdaestimates} are $\sim 2$.  Finally, the superpartner counterparts of the diagrams discussed in Table \ref{Tab:color5} will also correct the final values of the generated masses. Owing to these effects, there is some uncertainty in the above estimates and they may receive $\OrderOne$ corrections. However, we emphasize that these $\OrderOne$ differences in $\lambda$ will not affect the hierarchy between the generated masses. The hierarchy is always present in this model due to the repeated use of the same spurion to generate fermion masses (see section \ref{Spurions}). Since the use of the spurion involves a loop, the size of the hierarchy is determined by loop factors $\sim \l \loopfactor \r$. 

In the absence of $SU(5)$ breaking, leptons and down quarks will have equal masses. However, the familiar $SU(5)$ GUT relation $y_{b} = y_{\tau}$ is difficult to fit with the observed masses either in the standard model or  the MSSM   \cite{Xing:2007fb, Antusch:2009gu, Altmannshofer:2008vr}. This tension persists in split SUSY \cite{Giudice:2004tc}, where $\frac{y_\tau}{y_b} \approx 1.7$ when the SUSY breaking scale $\sim \Mgut$. In the domino mechanism, the generated quarks and lepton masses are not $SU(5)$ invariant since  they inherit the $SU(5)$ breaking in the higgs sector. Doublet-triplet splittings in the  $B_{\mu}$ and higgs scalar mass terms naturally induce $SU(5)$ violation in the generated masses. For example, when $\doubletbmu \ll \tripletbmu$, the color factors for the diagrams (see Figure  \ref{fig-tripletbtau}) that generate the $b$ and $\tau$ masses are $2$ and $3$ respectively (see Table \ref{Tab:color5}). Consequently, in this model, we naturally generate 
\begin{equation}
\frac{y_{\tau}}{y_b} = \frac{3}{2}
\end{equation}
which is close to the ratio  $\frac{y_\tau}{y_b} \approx 1.7$ expected in split SUSY \cite{Giudice:2004tc}. 

The $s$ mass measurement has a relatively large error. Owing to this error, the ratio $\frac{y_{\mu}}{y_s}$  lies between $3.5 \lessapprox \frac{y_\mu}{y_s} \lessapprox 6.6$ \cite{Xing:2007fb}, which is a significant deviation away from the  $SU(5)$ invariant relation $y_{\mu} = y_s$. This $SU(5)$ violation can also be accommodated in this model through doublet triplet splittings in the higgs sector. The requirement $\doubletbmu \ll \tripletbmu$ implies that the color factors for the diagrams that generate the $\mu$ and $s$ masses are in the ratio $3:2$. The value of these diagrams are further affected by the masses of the doublet and triplet components of $h_d$. Splitting in these masses alter the $\log$ factors that appear in \eqref{looplogfactor}. Moderate $\OrderOne$ differences in the doublet and triplet masses can easily split $y_{\mu}$ and $y_s$ further and accommodate the observed $SU(5)$ violation in the $\mu$ and $s$ masses. The observed $SU(5)$ violations in the down quark and lepton sectors are thus incorporated in the domino mechanism through  simple GUT breaking effects such as doublet-triplet splitting.  We note that while this radiative mechanism can naturally incorporate $\OrderOne$ $SU(5)$ violations, it does not naturally generate larger violations of $SU(5)$ invariance. 

The radiative contribution from the domino mechanism to the masses of the first generation are at the right order of magnitude for  $\lambda_{1j} \approx 1$.  These masses are also affected by the presence of Planck-suppressed operators (see section \ref{PlanckSlop}) and $\hdvev$ (see section \ref{FinalMass}) operators. As discussed in section \ref{FinalMass}, these operators are the sources of CP violation for the model since $U(1)$ rotations can be used to make all entries of the spurion $\lambda$ real.  The contributions from the operators 
\begin{eqnarray} 
\frac{\sigma^2 H_u 10_i 10_j}{M_{pl}^2} \qquad \text{and} \qquad \frac{\sigma^{\dagger} \HuDagger 10_i \FiveBar_j}{M_{pl}^2}
\end{eqnarray}
of section \ref{PlanckSlop} are comparable to the first generation masses when 
\begin{eqnarray}
\l\frac{\sigmavev}{M_{pl}}\r^2 \sim   \frac{F_{\sigma}}{M_{pl}^2} \sim 10^{-5}
\end{eqnarray} 
These conditions are satisfied when everything is around the GUT scale
\begin{eqnarray}
 \sigmavev \sim \Mgut \approx 10^{16} \text{ GeV} \qquad \text{and} \qquad  F_{\sigma} \sim \Mgut^2 \approx \l 10^{16} \text{ GeV} \r^2  .
 \end{eqnarray}
A vev $\hdvev$ for $h_d$ directly causes down quark and lepton masses through the operators $\lambdac 10_i \FiveBar_j$. In this model, $\hdvev$ must be tuned to be smaller than $\huvev$ (see section \ref{DownMasses}). If $\hdvev \approx 10^{-5} \huvev$, the contributions from this vev are comparable to the first generation down quark and lepton masses. 

Numerically, we find that the observed mixing angles between the first generation and other generations are naturally produced with planck suppressed operators discussed above. Furthermore, $\OrderOne$ phases introduced by these operators are sufficient to generate the observed CKM phase. For example, the Jarlskog invariant $J \sim 3 \times 10^{-5}$ \cite{PDG} is reproduced with a coefficient $\approx e^{-2.5 i}$ for the operator $\sigma^{\dagger} \HuDagger 10_1 \FiveBar_2$ and real coefficients for all other operators. 

The domino mechanism naturally reproduces the observed hierarchy in the quark and lepton masses with $\OrderOne$ yukawas $\lambdac$ ranging between $1$ and $3$. GUT breaking effects, quark mixing angles and the CKM phase are also easily incorporated into this framework. 

\subsection{Model with 45's}
\label{Model45Numbers}

The fermion mass hierarchy is generated in this model from the superpotential (see section \ref{Model45})
\begin{equation}
\mathcal{W} \supset \frac{y_3}{8} \, 10_3 \, 10_3 \, H_u \, +  \, \frac{x_1}{4} \, 10_{N_2} \, 10_3 \, H_u \,+  \, \frac{x_2}{4} \, 10_{N_2}\, 10_3 \, \phi \,  +  \, \frac{\lambdac}{2} \, 10_i \, \FiveBar_j \, \phibar 
\end{equation}
where $y_3$ is defined in equation \eqref{top45eq1} to be 
\begin{eqnarray}
y_3 & = & \l \frac{a_{11} \sigmavev}{M_1} \r^2 
\nonumber \\
\end{eqnarray}
$x_1$ and $x_2$ are obtained from equation \eqref{45topcouplings} and are given by 
\begin{eqnarray}
x_1 & = & \l \frac{a_{12} \sigmavev}{M_1}\r 
\nonumber \\
x_2 & = & \l \frac{b_{12} \sigmavev}{M_1}\r .
\nonumber 
\end{eqnarray}
The SUSY breaking $B_{\mu}$ term that communicates the chiral symmetry breaking to the $10_i \otimes \FiveBar_j$ space is 
\begin{equation}
\mathcal{L}_{\text{\sout{SUSY}}} \supset  \frac{B_{\mu}}{2} \, \tilde{\phi} \, \tilde{\phibar}
\end{equation}

The color factors for the diagrams that generate the fermion mass hierarchy in this model are summarized in Table \ref{Tab:color45}. The divergence structure of these diagrams are identical to those of the diagrams discussed in section \ref{PureSplitNumbers}. We diagonalize the generated yukawa matrices through the procedure outlined in section \ref{PureSplitNumbers}.  Taking
\begin{equation}
\begin{array}{lll} 
y_3 = \l \frac{a_{11} \sigmavev}{M_1}\r^2 \approx 1 \qquad & x_1 = \l \frac{a_{12} \sigmavev}{M_1}\r \approx 1 \qquad & x_2 = \l \frac{b_{12} \sigmavev}{M_1}\r \approx 1 \\ \\
\frac{B_{\mu}}{m^{2}_{\phibar}} \approx 0.05 \qquad & \log \l \frac{\Lambda^2}{m^{2}_{\phibar}}\r \approx 4 & \\ \\
\end{array}
\label{lambdaestimates45}
\end{equation}
we can fit the masses and mixing angles (from eqn.~\eqref{actualdata}) of the second and third generation with 
\begin{equation}
\begin{array}{lll}
\lambda_{33} \approx 0.8 \qquad & \lambda_{23} \approx 1.6 \qquad & \lambda_{22} \approx 2.5
\end{array}
\end{equation}

The masses and mixing angles of the first generation can be accommodated in this model with $\lambda_{1i} \approx 1$. But,  as in the case of the minimal model, these masses also receive contributions from Planck suppressed operators (see section \ref{Planck45}).  The observed CKM phase can be reproduced in this model with $\OrderOne$ phases introduced by these operators. 

The $45$ of $SU(5)$ can be decomposed into 7 standard model multiplets. $\OrderOne$ differences in the masses of these multiplets will result in $\OrderOne$ violations of the $SU(5)$ invariant GUT relations in the down quark and lepton sectors (see section \ref{PureSplitNumbers}).  Since there are three $SU(5)$ violating down quark and lepton masses and 6 mass splittings (between the 7 multiplets), we can accommodate the observed $\OrderOne$ $SU(5)$ violations. 

\begin{table}
\begin{center}
\begin{tabular}{|c|c|c|}
\hline
Description &  Figure   &   Color Factor  \\
\hline
Up sector, two loop& \ref{fig-2loopup} & $\frac{21}{4}$ \\
\hline
Down sector, two loop& \ref{fig-2loopdown}& $\frac{9}{8}$ \\
\hline
$10$ wavefunction renormalization& \ref{fig-wavefunction10} & $\frac{9}{2}$\\
\hline
$\FiveBar$ wavefunction renormalization& \ref{fig-wavefunction5} & 9\\
\hline
 One loop $b-\tau$ mass & \ref{fig-1loopbtau} & $-\frac{3}{2}$\\
\hline
3 loop contribution to $s-\mu$ & \ref{fig-3loop} & $\frac{243}{16}$\\
\hline
2 loop mixing & \ref{fig-2loopnew} & $\frac{123}{4}$\\
\hline
\end{tabular}
\end{center}
\caption[Color Factors]{\label{Tab:color45}  Color factors for the diagrams of the model with 45's discussed in section \ref{Model45}. These diagrams are the analogues of the diagrams discussed in section \ref{Model5}, with the appropriate replacements $H_u \rightarrow \phi$ and $H_d  \rightarrow \phibar$.  }
\end{table}

\section{Predictions}

In this Section we discuss further experimental signatures of our flavor model beyond the observed patterns in the quark and lepton Yukawa couplings.  Although the flavor structure is generated at the GUT scale it is not impossible to detect.  In our model this scale can be probed by late decaying particles, including the proton and the gluino.  Some of the signatures, notably proton decay, specifically probe the mechanism producing the flavor structure of the SM.

\subsection{Proton Decay}

Since the flavor structure of the SM is generated near the GUT scale, it is difficult to directly observe this mechanism in colliders.  However, proton decay provides a probe of such high scales.  In our model, Eqn. \eqref{Eqn: flavor superpot}, proton decay arises from the second term:
\begin{equation}
\label{eqn: p-decay coupling}
\mathcal{W} \supset \lambda_{ij} \, 10_i \, \FiveBar_j \, \phibar
\end{equation}
Since the $\lambda_{ij}$ are all $\OO(1)$ couplings, these decays are not Yukawa suppressed.  As we discussed, $\phibar$ can be either a $\bar{5}$ or a $\bar{45}$ of SU(5).  As an example, we will consider the case that $\phibar$ is a $\bar{5}$ (i.e. it is the down-type higgs, $H_d$) in detail.  In this case it is the color triplet component, $h^{(3)}_d$, that causes proton decay.  If $\phibar$ is a $\bar{45}$, three of its SM components cause proton decay, the $(3,1,-\frac{1}{3})$, $(3,3,-\frac{1}{3})$, and $(\bar{3},1,\frac{4}{3})$.  However the last two components only cause proton decay through the coupling of the $45$ to $10 \, 10$ and thus are quite Yukawa suppressed in our model \cite{Dorsner:2009cu}.  The proton decay predictions in this case are similar to the previous case though the order 1 numbers multiplying the various contributions may be different.

The $h^{(3)}_d$ scalar gives rise to the dimension 6 proton decay operator in the Kahler potential $K \supset Q L U^\dagger D^\dagger$.  The normal dimension 5 proton decay operators of the MSSM are suppressed by the large masses of the squarks and sleptons.  The dimension 6 proton decay operators mediated by X, Y gauge bosons are present, but the $h^{(3)}_d$ is likely to give the dominant contribution to the proton decay rate since the rate scales as the fourth power of the mediating particle's mass and the gauge contribution is suppressed by gauge couplings.  The consequences of X, Y mediated proton decay are not as specific to this model (e.g. \cite{Machacek:1979tx, Murayama:2001ur, Dorsner:2004xa}) and would not give as direct a window into the flavor structure so we will not discuss them further and instead assume the proton decay rate is dominated by $h^{(3)}_d$ exchange.

In this case, the predicted proton branching ratios are interestingly different from the predictions of other theories \cite{Dimopoulos:1981dw, Lucas:1996bc}.  Because $\lambda$ in Eqn. \eqref{eqn: p-decay coupling} is an arbitrary matrix with $\OO(1)$ entries, the decay rates are roughly equal into all the different possible two-body final states.  Notice that this includes the unusual decay modes $p \to \mu^+ \pi^0$ and $p \to e^+ K^0$.  The decay rates are given by s- and t-channel exchange of the $h^{(3)}_d$ scalar and are proportional to the following combinations of the couplings:
\begin{equation}
\begin{array}{rclcrcl}
\Gamma \left(p \to e^+ \pi^0 \right) & \propto & \lambda^4_{11} & \qquad & \Gamma \left(p \to \nu_e \pi^+ \right) & \propto & \lambda^4_{11} \\
\Gamma \left(p \to \mu^+ \pi^0 \right) & \propto & \lambda^2_{11} \, \lambda^2_{12} & \qquad & \Gamma \left(p \to \nu_\mu \pi^+ \right) & \propto & \lambda^2_{11} \, \lambda^2_{12} \\
\Gamma \left(p \to e^+ K^0 \right) & \propto & \lambda^2_{11} \, \lambda^2_{12} & \qquad & \Gamma \left(p \to \nu_e K^+ \right) & \propto & \lambda^2_{11} \, \lambda^2_{12} \\
\Gamma \left(p \to \mu^+ K^0 \right) & \propto & \lambda^4_{12} & \qquad & \Gamma \left(p \to \nu_\mu K^+ \right) & \propto & \left( \lambda^2_{12} + \lambda_{11} \, \lambda_{22} \right)^2
\end{array}
\end{equation}
And the neutron decay modes (which can be measured almost as well) are
\begin{equation}
\begin{array}{rclcrcl}
\Gamma \left(n \to e^+ \pi^- \right) & \propto & \lambda^4_{11} & \qquad & \Gamma \left(n \to \nu_e \pi^0 \right) & \propto & \lambda^4_{11} \\
\Gamma \left(n \to \mu^+ \pi^- \right) & \propto & \lambda^2_{11} \, \lambda^2_{12} & \qquad & \Gamma \left(n \to \nu_\mu \pi^0 \right) & \propto & \lambda^2_{11} \, \lambda^2_{12} \\
\Gamma \left(n \to e^+ K^- \right) & \propto & 0 & \qquad & \Gamma \left(n \to \nu_e K^0 \right) & \propto & \lambda^2_{11} \, \lambda^2_{12} \\
\Gamma \left(n \to \mu^+ K^- \right) & \propto & 0 & \qquad & \Gamma \left(n \to \nu_\mu K^0 \right) & \propto & \left( \lambda^2_{12} + \lambda_{11} \, \lambda_{22} \right)^2
\end{array}
\end{equation}
where the zeros just mean the process does not happen at leading order.  Note that there are different $\OO(1)$ numbers in front of each of these terms coming from combinatoric factors whose measurement would help confirm the underlying model giving rise to the decay.  Of course, in practice it is impossible to distinguish the different flavors of neutrino coming from proton decay so the rates for $\nu_e$ and $\nu_\mu$ must be added.  Not only does the presence of many decay modes enhance the observability of the signal, it also allows many independent measurements of the parameters and cross-checks of this framework.  It would be a remarkable discovery to observe the mechanism that is responsible for the flavor structure of the SM in the various branching ratios of proton decay.

The rates are all roughly
\begin{equation}
\Gamma \sim \frac{1}{8 \pi} \lambda^4 \frac{m^5_p}{M^4_h} \approx \frac{1}{10^{35} \text{ yr}} \,  \lambda^4 \left( \frac{2 \times 10^{16} \, \GeV}{M_h} \right)^4
\end{equation}
where $M_h$ is the mass of the $h^{(3)}_d$ scalar.  Taking into account the actual hadronic matrix elements may if anything make this decay rate slightly faster \cite{Murayama:2001ur}.  Excitingly, the next generation of proposed proton decay experiments, including at DUSEL and Hyper-K, should be sensitive to lifetimes up to about $10^{35}$ yr \cite{Raby:2008pd}.  Thus it seems likely that at least some of our proton decay channels should be accessible to the next generation of proton decay experiments.  This is an important prediction of our model since a proton decay signal would directly probe the couplings that give rise to the quark and lepton masses and mixings in the SM.

\subsection{Strong CP and the Axion}

Our mechanism explains flavor by forbidding the Yukawa couplings with a $U(1)$ symmetry and then generating them when that symmetry is spontaneously broken.  This necessarily leads to a Goldstone boson, $a$.  Further this $U(1)$ symmetry has a mixed anomaly with the $SU(3)_C$ gauge group, as would be generally expected of a new global symmetry (for the $U(1)$ charges see Table \ref{Tab:charges} for the model of Section \ref{Model5}, and Table \ref{Tab:charges45} for the model of Section \ref{Model45}).  Then in the low energy theory, $a$ picks up a coupling in the Lagrangian from this anomaly $\propto \frac{a}{f} G \widetilde{G}$ and a small mass when QCD condenses $\propto \frac{\Lambda_\text{QCD}^2}{f}$.  Thus $a$ is the QCD axion, and our approach to flavor necessarily also solves the strong CP problem.

The color anomaly of our $U(1)_H$ is generated by the quarks of the SM, $10_i$ and $\FiveBar_i$, as well as the new colored particles in the $10_N$.  $H_u$ and $H_d$ are irrelevant since they are vector-like under $U(1)_H$.  In this way we have a combination of DFSZ \cite{Dine:1981rt, Zhitnitsky:1980tq} and KSVZ \cite{Kim:1979if, Shifman:1979if} style axions.  The axion is generated when $\sigma$ gets a vev, breaking the $U(1)_H$ symmetry.  The scale of symmetry breaking sets the scale suppressing the couplings of the axion, $f \sim \langle \sigma \rangle \sim \Mgut$.  At the weak scale there is only one Higgs, $H_u$, which gets a vev and whose Goldstone bosons are eaten by the W and Z.  Though these Goldstones mix slightly with the Goldstone from $\sigma$, the axion $a$ is dominated by that from $\sigma$ and hence its couplings are suppressed by the scale $f \sim \Mgut$ as in the DFSZ model.

Besides the usual split SUSY spectrum and the axion, the only other particle we predict at low energies is the axino, $\widetilde{a}$, the fermionic superpartner of the axion.  In the absence of SUSY breaking the axino would be as light as the axion.  When SUSY is broken, the saxion, the other scalar superpartner of the axino, usually gets a mass of order the squark and slepton masses, which for us is near the GUT scale.  The axino however can get a mass far below the SUSY breaking scale \cite{Chun:1992zk}.  Just how far is, in general, dependent on the details of the SUSY breaking sector.  Since our goal is to write down an effective theory describing flavor and give its general predictions, we will not attempt to write down a specific model for the axino mass.  However, we do wish to know whether we expect the axino to get a mass of order the scalar superpartners, $\sim \Mgut$ or the much lighter scale of the gauginos and higgsinos, $\sim$ TeV.  The effective axion and axino couplings arise from integrating out the $10_i$, $\FiveBar_i$, and $10_N$ fields as described above and give rise to the superpotential operators
\begin{equation}
\label{eqn: axino coupling}
\mathcal{W} \propto \frac{\alpha_s}{4 \pi} \frac{S}{f} \, G_\alpha G^\alpha + \frac{\alpha_\text{EM}}{4 \pi} \frac{S}{f} \, F_\alpha F^\alpha
\end{equation}
where $G$ is the QCD and $F$ is the EM field strength operator.  The imaginary part of the scalar component of $S$ contains the axion $a$, and the fermionic component of $S$ contains the axino.  In order to allow the normal kinetic terms, $W_\alpha$ must have R-charge +1 under the R-symmetry which keeps the gauginos and higgsinos light (at the TeV scale) in split SUSY.  Therefore $S$ has R-charge 0 and so the axino mass, which is the F-term of $S \, S$, is forbidden by the R-symmetry.  This R-symmetry is only broken at the TeV scale to give the gauginos and higgsinos their masses, therefore the axino mass is controlled by the TeV scale and not the GUT scale.  Thus it is reasonable to expect the axino to be much lighter than the TeV scale, as in normal supersymmetric models \cite{Chun:1992zk}.

Such axions with $f \sim \Mgut$ and mass $\sim 10^{-9}$ eV could make up the dark matter of the universe.  In fact the initial displacement angle has to be tuned to avoid producing too much dark matter.  They are difficult, though perhaps not impossible \cite{Arvanitaki:2009fg}, to detect since all couplings to SM particles, $G \widetilde{G}$, and $F_\text{EM} \widetilde{F}_\text{EM}$ are suppressed by the scale $f$.  The axino could also be a component of the dark matter, though since its couplings are similar to the axion's, it may also be difficult to detect directly \cite{Covi:2001nw, Covi:2009pq}.

\subsection{Long-Lived Particles and Late Decays}

In our scenario the axino can easily be the true lightest supersymmetric partner (LSP).  In this case, all other superpartners will decay to it through the dimension 5 operators in Equation \eqref{eqn: axino coupling}.  In particular the next to lightest supersymmetric partner (NLSP), usually the lightest neutralino, which would otherwise have been a stable dark matter particle, will decay to the axino.  Thus the dark matter today will be made up of axions and axinos only.  Additionally, split SUSY generally has a long-lived gluino which can be cosmologically stable \cite{ArkaniHamed:2004fb}.  When the squark and slepton masses are around $10^{14}$ GeV as in our scenario, the gluino has a lifetime around the age of the universe.  Such long-lived colored particles are constrained by cosmic ray measurements as well as searches for anomalously heavy elements and strongly interacting dark matter and generically rule out any SUSY breaking scale (squark and slepton mass scale) higher than $\sim 10^{10} - 10^{12} \, \GeV$ \cite{Arvanitaki:2005fa}.  However, in our model this is not a problem because the gluino decays through the dimension 5 operator of Equation \eqref{eqn: axino coupling} to the axino.  This solution of the long-lived gluino problem in split SUSY applies whenever a QCD axion is added to the theory.

The lifetime for the decay of the gluino to the axino through the operators in Equation \eqref{eqn: axino coupling} is
\begin{equation}
\tau \sim 8 \pi \left( \frac{4 \pi}{\alpha_s} \right)^2 \frac{f^2}{m^3} = 2 \times 10^4 ~\s \left( \frac{\TeV}{m} \right)^3 \left( \frac{f}{10^{16} ~\GeV} \right)^2
\end{equation}
where $m$ is the mass of the gluino.  Since $\alpha_\text{EM} < \alpha_s$, the lifetime of the NLSP would be longer.  This lifetime has some variability depending on the exact values of the scale $f$ and the gluino mass $m$.  It is generally around the time of big bang nucleosynthesis (BBN).  Thus the gluino avoids all the late time constraints, and is only possibly constrained by BBN.  The relic gluino abundance is uncertain by several orders of magnitude due to nonperturbative effects and thus could easily avoid BBN constraints.  Interestingly, these lifetimes for the gluino and the NLSP are in the right range to explain the cosmological Lithium problems.  The measured primordial abundances of $\LiSix$ and $\LiSeven$ are significantly different from the predictions of standard BBN and in opposite directions from each other.  Both of these problems can be solved simultaneously by a long-lived particle decaying with a lifetime of $\sim 1000$ s and with roughly the abundance and hadronic branching ratio of either the gluino or the NLSP.  Thus the presence of the axino in our model not only solves the long-lived gluino problem of split SUSY but may also naturally solve the cosmological Lithium problems.  Such late-decays are discussed in more detail in \cite{Arvanitaki:2008hq}.

Importantly, the long-lived gluino could potentially be discovered at the LHC with a reach above 1 TeV through monojets, dijets, and charged tracks \cite{Hewett:2004nw, Fairbairn:2006gg}.  Although the NLSP will fly straight out of the LHC's detectors (assuming it is the lightest neutralino), some fraction of gluinos that are produced at the LHC will stop in the detectors.  Their out-of-time decays could give a dramatic signature of such a scenario and also a measurement of the gluino mass and lifetime \cite{Arvanitaki:2005nq}.  In our case the gluino dominantly decays two-body to a jet plus missing energy, which could distinguish this scenario from more standard split SUSY decays in which the gluino decays to two jets plus missing energy with a significant branching fraction.  In fact such a search has already been performed at the Tevatron \cite{Abazov:2007ht}.  Not only would this signal provide a unique confirmation of the particle physics solution of the cosmological Lithium problems, the measurement of the mass and lifetime would provide a measure of the axion decay constant $f$ and hence powerful evidence for a new scale in nature near the GUT scale.

\subsection{Higgs Mass}
\label{Sec: higgs mass}

Our model has a sharp prediction for the Higgs mass.  In Split SUSY with the SUSY breaking scale near the GUT scale, the Higgs mass is tightly constrained and essentially independent of the actual value of the scalar masses \cite{Arvanitaki:2004eu}.  The value of the Higgs mass is determined by RG evolving the Higgs quartic from the high scale.  In our minimal model where $\tan \beta$ is large, the Higgs is in the range roughly 148 GeV to 154 GeV \cite{Binger:2004nn}.  In the model with 45's the value of $\tan \beta$ can be lower so the Higgs mass can range down to roughly 140 GeV.  Most of the uncertainty arises from errors in the top mass and $\alpha_s$.  If these errors are reduced in the future, the Higgs mass prediction will become even sharper.  This range will be accessible at the LHC and may even be reached by the Tevatron.

\section{Conclusions}

The nontrivial structure of the Yukawa couplings is one of the few pieces of experimental evidence we have that may provide a hint to the nature of physics beyond the standard model.
Even the hypothesized existence of a gigantic landscape of vacua could not explain the structure of the Yukawas (at the very least of the heavier two generations).
Although many models of flavor exist, our model differs significantly from these in its structure, implications for flavor, and testable predictions.  Froggatt-Nielsen and many extra-dimensional models require different, ad hoc choices of charge or position for each different fermion in order to reproduce the observed structure in the Yukawa couplings.  Alternative extra-dimensional models rely on random, exponentially small overlaps between different fermions' wavefunctions to produce small Yukawa couplings.  While these frameworks explain the smallness of the Yukawa couplings and can accommodate their observed hierarchical structure, they could just as easily have accommodated any other structure.

By contrast, our model makes definite predictions for the hierarchies between the masses of successive generations and for their mixing angles.  Further, all generations are treated identically, there is no need to single out any particle (e.g. the top quark) by choice.  We have a new symmetry which forbids the Yukawas and when that is broken spontaneously, the Yukawa couplings are all naturally generated radiatively in the observed hierarchical patterns.  We follow the effective field theory philosophy of simply writing down every allowed term with $\OO(1)$ coefficients.  Because of the simple structure of the two couplings that collectively break all flavor symmetries, successive generations only receive mass at successive loop orders as in Figure \ref{Fig: dominopattern}.  This ``domino mechanism" produces the observed hierarchies in the Yukawa couplings from loop factors.  Interestingly, the models naturally produce small quark mixing angles but large neutrino mixing angles.  Note that we have generated the Yukawa couplings of the downs and leptons as $H^\dagger_u \, 10 \, \FiveBar$ instead of with $H_d$.  We have presented two models that illustrate this mechanism, the model of Section \ref{Model5} (summarized in Equations \eqref{UglyPotential} and \eqref{eqn: b mu term minimal model}), and the model of Section \ref{Model45} (summarized in Equations \eqref{Eqn:Potential45} and \eqref{Eqn:Potential45num2}).

Our models naturally preserve unification.  They are embedded in split SUSY where both the SUSY breaking scale and the scale of flavor (the breaking scale of our new symmetry) are at the GUT scale, which is the only scale in the model besides the weak scale.  Our mechanism gives novel SU(5) breaking relations in the Yukawa couplings.  For example, in our minimal model the $\tau$ to b mass ratio is predicted to be $\frac{3}{2}$ instead of 1, resolving the tension in many SUSY models.  It arises as the ratio of the number of colors that run inside the loop in Figure \ref{fig-tripletbtau}, which is the number of colors the fermion does not have (3 for the $\tau$ and 2 for the b).  A ratio larger than 1 is easily produced in our second model as well.  In both models the $\mu$ to s mass ratio also receives SU(5) breaking contributions which can easily split the masses by an $\OO(1)$ factor.  Such mass ratios arise from the SU(5) breaking in the masses of the GUT scale particles which are running in the loops that generate the Yukawas.  This is an interesting illustration that the common problems of SU(5) Yukawa unification in GUTs may not indicate a problem with the supersymmetric framework but may instead arise naturally from a specific model of flavor.

Current experimental flavor constraints indicate that the Yukawas may not be generated near the weak scale.  Although the SM flavor structure could be generated at any new intermediate scale, the GUT scale is already a well motivated scale and so seems a likely place for flavor to be generated.  This case seems pessimistic since high-scale flavor models are often difficult to test experimentally.  Remarkably though, our models have several potentially testable predictions.  Importantly, these models give novel proton decay predictions which may well be observable at the next generation of proton decay experiments.  Measurements of the various proton decay modes and branching ratios would be a direct probe of the couplings that generate the flavor structure of the standard model.  Our mechanism for explaining flavor also necessarily solves the strong CP problem by producing an axion.  The Yukawas are explained by forbidding them with a symmetry and generating them when that symmetry is broken.  That symmetry has a mixed anomaly with QCD and therefore gives rise to an invisible axion when it is broken at the GUT scale.  Additionally it necessarily implies an axino which can easily be the lightest superpartner.  Thus all superpartners, including the long-lived gluino of split SUSY will decay to the axino, resolving the problems with high scale split SUSY.  Such decays can have observable effects on the light element abundances produced during BBN and may even explain the Lithium problems.  Further, if the gluino is produced at the LHC, it can stop and decay out of time in the detectors, giving a dramatic signature of such a scenario and further evidence for the GUT scale.  Finally the Higgs mass is predicted to lie within a range from around 140 GeV to 154 GeV, which should be tested soon.

Although our specific models rely on radiative generation of the Yukawa couplings, the basic mechanism may be more generally applicable.  The hierarchies between the different generations arise just from the simple flavor structure of our two flavor symmetry breaking couplings, the vector and the matrix.  The pattern in the Yukawas is generated from the necessity of using these spurions repeatedly until all flavor directions are generated, as in Eqns.~\eqref{Eqn: up-yukawa spurions} and \eqref{Eqn: down-yukawa spurions}.  So long as each use of a spurion introduces a small factor in the coupling, the general hierarchical pattern will be reproduced.  Our radiative model allows this spurionic factor to be rigorously computed and the numerical values work well.  But there may be other implementations of this same idea that also give interesting results.  The simplicity of the two spurions may lead, for example, to novel ways to embed the flavor structure of the Standard Model in string constructions.

\section*{Acknowledgments}
We would like to thank Nima Arkani-Hamed, Asimina Arvanitaki, Nathaniel Craig, Savas Dimopoulos, Ilja Dorsner, Sergei Dubovsky,  Daniel Green, Lawrence Hall, Roni Harnik, John March-Russell, Stuart Raby, Scott Thomas, and Jay Wacker for useful discussions. We thank the Institute for Advanced Study and the Oxford theory group for their hospitality during completion of this work.  This work was partially supported by the Department of Energy under contract number DE-AC02-76SF00515.  PWG was partially supported by NSF grant PHY-0503584.

\end{document}